\documentclass{aa}  

\usepackage{graphicx}
\usepackage{txfonts}
\usepackage{xcolor}
\usepackage{soul}
\usepackage{comment}
\usepackage{multicol}
\usepackage{geometry}
\usepackage{hyperref}
\usepackage{longtable,booktabs}
\usepackage{stfloats}
\begin{document}

   \title{Frequency evolution of pulsar emission: }

   \subtitle{Further evidence for fan beam model}

   \author{P. Jaroenjittichai\inst{1}\fnmsep\thanks{phrudth@gmail.com}
        \and
        S.~Johnston\inst{2}
        \and
        S.~Dai\inst{2}
        \and
        M.~Kerr\inst{3} \and
        M.~E.~Lower\inst{4} \and
        R.~N.~Manchester \inst{2} \and
        L.~S.~Oswald\inst{5}\and
        R.~M.~Shannon\inst{4}\fnmsep\inst{6}\and
        C.~Sobey\inst{7}\fnmsep\inst{8}\and
        P.~Weltevrede\inst{9}
        }

   \institute{National Astronomical Research Institute of Thailand (Public Organization), 260 M.4, Donkaew, Maerim, Chiang Mai, 50180, Thailand 
         \and
             Australia Telescope National Facility, CSIRO, Space and Astronomy, PO Box 76, Epping, NSW 1710, Australia
             \and
Space Science Division, Naval Research Laboratory, Washington, DC 20375, USA
\and
Centre for Astrophysics and Supercomputing, Swinburne University of Technology, PO Box 218, Hawthorn, VIC 3122, Australia
\and
School of Physics and Astronomy, University of Southampton, Southampton SO17 1BJ, UK
\and
OzGrav: The ARC Centre of Excellence for Gravitational-wave Discovery
\and
SKAO, ARRC Building, 26 Dick Perry Avenue, Kensington, WA 6151, Australia
\and
CSIRO, Space and Astronomy, PO Box 1130 Bentley, WA 6102, Australia
\and
Jodrell Bank Centre for Astrophysics, The University of Manchester, Alan Turing Building, Manchester, M13 9PL, United Kingdom
             }  

   \date{Received ; accepted }

 
  \abstract

{We explore frequency-dependent changes in pulsar radio emission by analyzing their profile widths and emission heights, assessing whether the simple radius-to-frequency mapping (RFM) or the fan beam model can describe the data.}
{Using wideband (704–4032 MHz) Murriyang (Parkes) observations of over 100 pulsars, we measured profile widths at multiple intensity levels, fitted Gaussian components, and used aberration–retardation effects to estimate emission altitudes. We compared trends in width evolution and emission height with a fan beam model.}
{Similar to other recent studies, we find that while many pulsars show profiles narrowing with increasing frequency, a substantial fraction show the reverse. The Gaussian decomposition of the profiles reveals that the peak locations of the components vary little with frequency. However, the component widths do, in general, narrow with increasing frequency. This argues that propagation effects are responsible for the width evolution of the profiles rather than emission height. Overall, the evolution of the emission height with frequency is unclear, and clouded by the assumptions in the model. Spin-down luminosity correlates weakly with profile narrowing but not with emission height.}
{The classic picture where pulsars emit at a single emission height which decreases with increasing observing frequency cannot explain the diversity in behavior observed here. Instead, pulsar beams likely originate from extended regions at multiple altitudes, with fan-beam or patchy structures dominating their frequency evolution. Future models must incorporate realistic plasma physics and multi-altitude emission to capture the range of pulsar behaviors.}


   \maketitle
%
\section{Introduction}
It is a well-established idea that the profiles of radio pulsars are narrower at higher radio frequencies (\citealt{cor78}). This trend is exemplified by pulsars such as PSR~B1133+16, where the two components of its profile move closer together as the observing frequency increases (\citealt{hsh+12}). Empirically, \cite{tho91} quantified the frequency evolution of the component separations with a power-law. We generalized their relation as follows:
\begin{equation}
    W_{x} = A_{x} \nu^{\mu_{x}} + W_{x,0},
    \label{eq:thorsett}
\end{equation}
where \(W_{x}\) represents the profile width at $x\%$ of the peak intensity, \(\nu\) is the observing frequency, \(\mu_{x}\) characterizes the frequency scaling, and \(W_{x,0}\) represents the asymptotic width. 
Multi-frequency data for tens pulsars analyzed in the 1990s can be fit by the Thorsett relation, seemingly confirming the general narrowing of profile width at high frequencies (e.g. \citealt{xkj+96,kg98,mr02}). The frequency-dependent profile evolution was called ``radius-to-frequency mapping (RFM)'' in the literature and is interpreted as meaning that high frequency radio emission arises close to the neutron star surface, whereas low frequency originates higher in the magnetosphere. This is due to the fact that the emission cone opening angle increases with altitude.

However, with a larger sample of pulsars available by 2010, it became obvious that not all pulsars conform to this idea. \citet{cw14} analyzed 150 pulsars and found that $\sim$54\% showed a significant profile narrowing at high $\nu$, but a sizable fraction ($\sim$19\%) exhibited broadening at high frequencies, with the rest showing little change. Similarly, \cite{pkj+21} investigated profile width evolution with frequency for 762 pulsars, revealing that many pulsars exhibit width broadening at higher frequencies, contrary to expectations. Such broadening trends cannot be explained by the simplistic RFM picture.
\begin{figure*}
        \centering
    \includegraphics[width=0.3\linewidth]{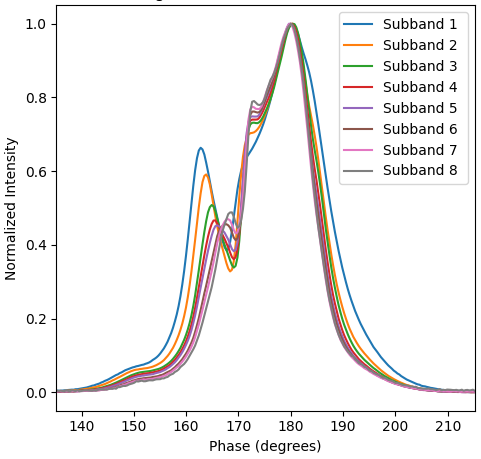}
    \includegraphics[width=0.3\linewidth]{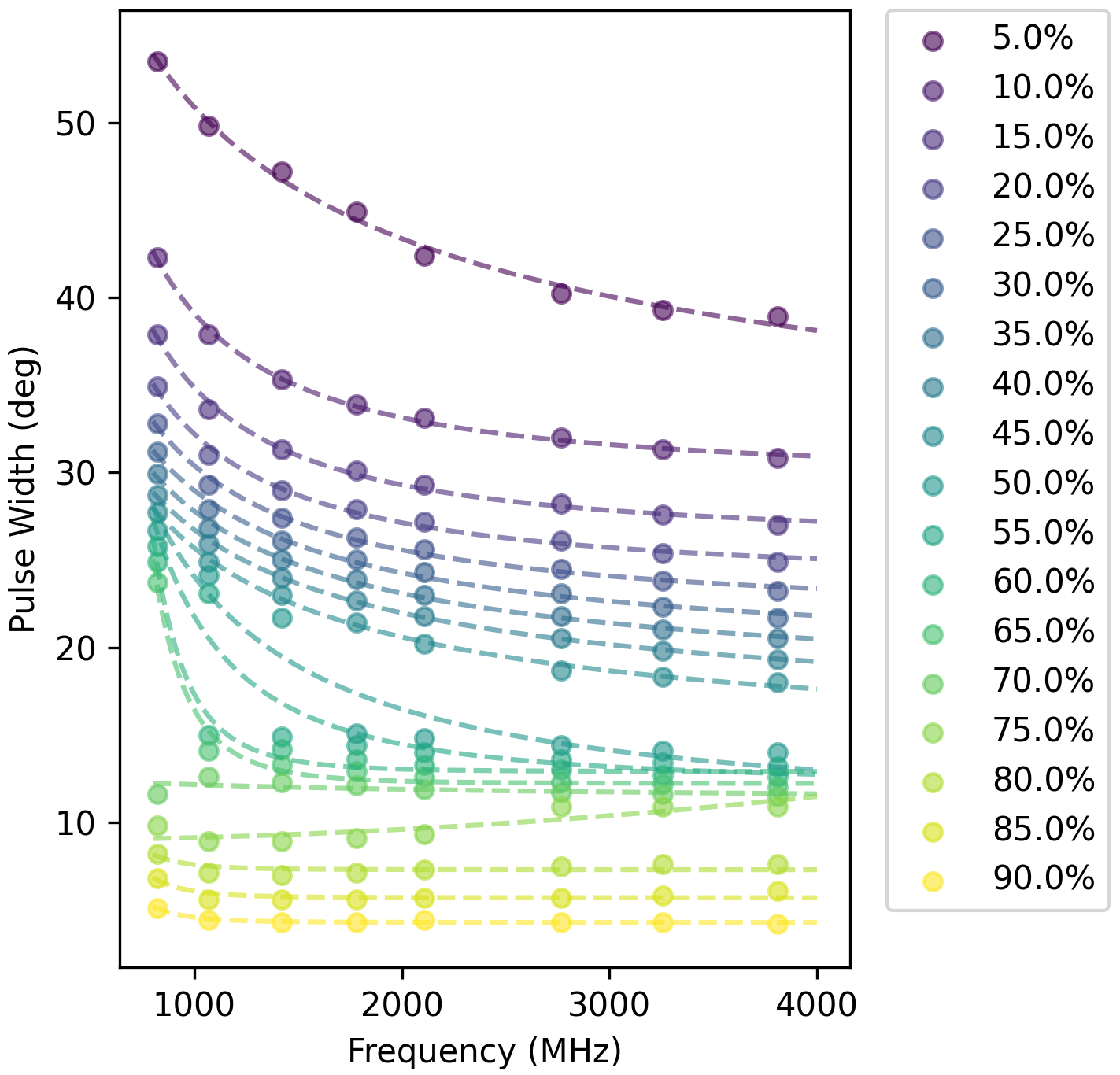}
    \includegraphics[width=0.31\linewidth]{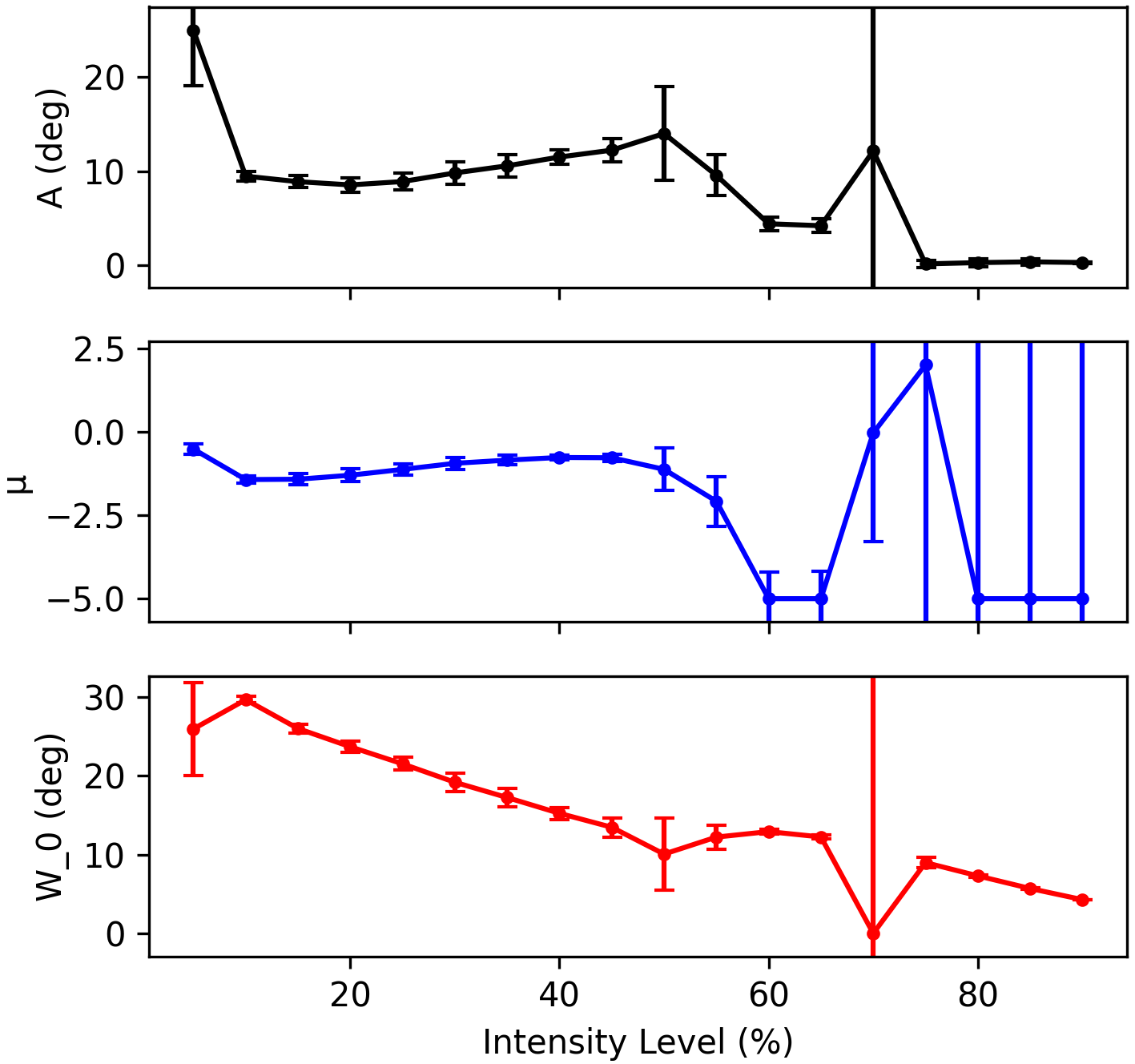}
    \caption{Profile width evolution of PSR~J0738--4042. The figure on the left shows the stack of normalized profiles, with the profile peak placed at 180\degr\ for each sub-band. ($Middle$) The profile width has been measured at varying intensity levels, shown from dark to light colors, in relation to frequency. The dashed lines represent the best-fit solutions of Eqn.~\ref{eq:thorsett}. The resulting parameters as a function of profile intensity level is shown in the right panel. The black, blue, and red represent the amplitude, $A$, the power index, $\mu$, and the y-offset, $W_0$, respectively.
    }
    \label{fig:0738}
\end{figure*}
\begin{figure}
        \centering
    \includegraphics[width=1\linewidth]{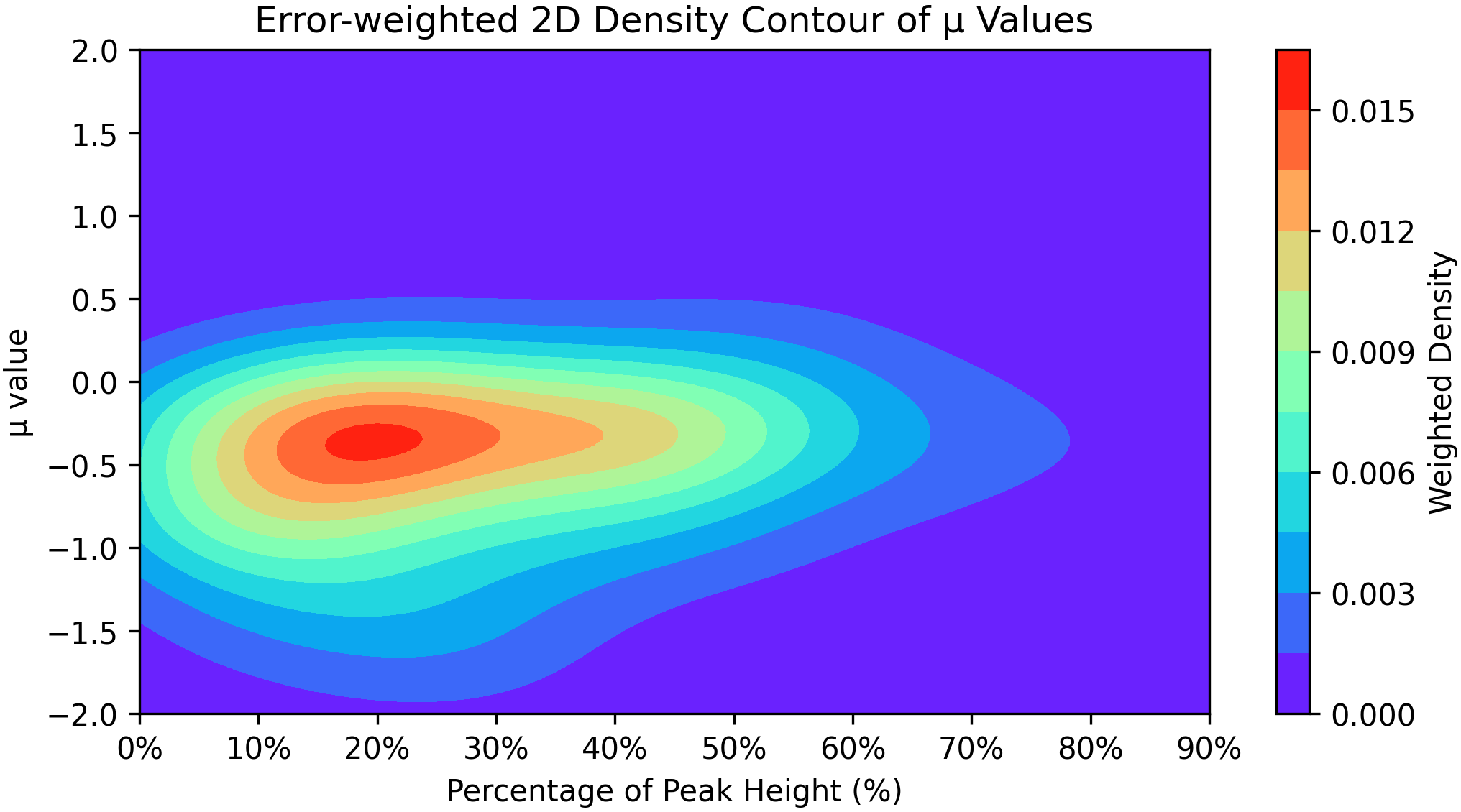}
    \caption{Distribution of frequency evolution index ($\mu$) of profile width measured at different profile intensity level ($W_{\rm x}$) for 144 pulsars. The 2D contour map shows the counts weighted by the uncertainty of each data point.}
    \label{fig:widthhistoweight}
\end{figure}

An alternative fan beam model has emerged to address these issues (e.g. \citealt{drd10,dr15,Wang2014,dp15}). Unlike the traditional nested cone model, the fan beam model explains profiles as arising from structured, azimuthally elongated beams rather than symmetric, concentric rings. The emission consists of elongated streams or fan-like beams of radiation from plasma flowing along specific magnetic field lines, creating a pattern of emission that resembles spokes on a wheel when viewed along the pulsar’s magnetic axis (see e.g. Fig.~2 in \citealt{dr15}).

\cite{dp15} argued that the observed peak separation ratio ($R_W$) in pulsar profiles is not consistent with the conal model but is well explained by the fan beam model. They analyzed pulsar profiles with four and five components and measured the separations between these components. They found that the conal model predicts a narrow, sharply peaked distribution of $R_W$, while the observed data show a broader, flatter distribution. The fan beam model, on the other hand, naturally produces the observed distribution. \cite{okj19} studied PSR~B1133+16, the pulsar largely responsible for the RFM model, using single-pulse observations at multiple radio frequencies. By comparing the observed frequency-dependent profile broadening with simulations, they demonstrated that profile widening at lower frequencies is more consistent with azimuthally elongated emission patches rather than simple nested cones. The fan beam model therefore serves as an alternative for statistical distributions of peak separations, frequency-dependent broadening, and single-pulse characteristics.

In this paper we measure profile width and emission height evolution across frequencies from 704 MHz to 4.032 GHz. Section 2 describes our dataset from Murriyang observations. Section 3 outlines our methodology. Section 4 presents our results on profile characteristics and emission heights. Section 5 discusses these findings and proposes a simple fan beam model. Section 6 summarizes our conclusions about pulsar emission geometry.

\section{Dataset}
The pulsars for this project were observed with Murriyang, CSIRO's Parkes radio telescope, under the auspices of project ID P574. The P574 project in its current form has been observing some 250 pulsars per month since late 2018 \citep{jsd+21}. Observations are made with the Ultra-Wideband Low (UWL) receiver \citep{hmd+20} operating in the frequency range 704 to 4032~MHz.  The data reduction is identical to that described in \cite{sjd+21} and also used by \cite{ojk+23}. In brief, the dataset we use here consists of flux and polarization calibrated profiles, with 1024 phase bins across the pulse period. The bandwidth of the UWL is divided into eight subbands, with center frequencies 822, 1070, 1420, 1780, 2107, 2766, 3258 and 3810 MHz.

Of the 250 pulsars regularly observed as part of P574 we removed pulsars with low flux density and those with clear signs of scatter broadening in the lower frequencies. This leaves a total of 157 pulsars. They are listed in Table 1 and Table 2 in the Appendix. 

\section{Methodology}
\subsection{Profile widths}
We employed three different methods to determine profile widths. First, the profile width was based on the percentage of the peak intensity. The profile width of each subband was  measured at 18 intensity levels, ranging from 5\% to 90\%, of the profile peak ($W_{\rm x}$). 
Linear interpolation was employed to determine profile edges at specified intensity thresholds. The profile width at each frequency was fitted with Eqn.~\ref{eq:thorsett}. The curve fitting process utilizes {\tt scipy.optimize.curve\_fit} with initial parameters $A$ = 100 for the amplitude coefficient, $\mu$ = 0.0 for the frequency scaling index, and $W = 5^{\circ}$ for the asymptotic profile width, along with boundary conditions constraining A between 0 and 2000, $\mu$ between $-5$ and 5, and $W$ between 0\degr\ and 100\degr. The implementation requires a minimum of four valid data points for each intensity level to ensure reliable fitting results. The fitted parameters and their uncertainties were obtained from the covariance matrix output from the fitting procedure.

As we show in sections 4 and 5 this methodology is flawed when examining the frequency evolution of the profiles. We therefore adopted a different approach which requires multi-Gaussian component fits to the profiles (e.g. \citealt{kwj+94}) at each subband. In the second method, we used the maximum component separation that we denote $W_{\rm comp.sep}$.
We first created a template profile by averaging the data in frequency within each subband and then fitted the profiles with multiple Gaussian components using a least-squares optimization approach. The fitting process iteratively added Gaussian components until the residuals fell below the off-pulse noise level. The parameters of this template fit were then used as initial conditions to analyze each frequency subband independently, with bounds set to ±20\% of the template values. The maximum component separation ($W_{\rm comp.sep}$) of each subband were then fitted with the same curve-fitting routine {\tt scipy.optimize.curve\_fit} as in the first method. The separation was measured from the center of the component.

Third, in contrast to the approach employed in the second method, we have chosen to use the width of each individual Gaussian component, denoted as ($W_{\rm comp.width}$). The values of $W_{\rm comp.width}$ for all components were fitted according to Eqn.~\ref{eq:thorsett}, using a curve fitting routine similar to the previous methods.

These latter two methods are also not without their caveats. Components are almost certainly not Gaussians, and Gaussian fitting can return non-unique solutions. However in the light of any obvious alternative and with the large sample of pulsars under consideration here, these methods appear to be the most robust way to tackle the problem.

\subsection{Emission height}
We estimated the pulsar emission height using two different methods. The first is a geometric method. The geometric emission height ($r_{\rm 90}$) correlates with the half-opening angle of the emission beam ($\rho$) and the star's rotation period ($P$). This correlation stems from the pulsar's dipolar magnetic field structure, which leads to the divergence of field lines from the star's surface. In this case $\rho = \sqrt{9 \pi r_{\rm 90}/ 2Pc}$, where $c$ is the speed of light. In turn, $\rho$ can be deduced from the observed profile width,$W$, the angle between the magnetic and rotation axes, $\alpha$, and the angle between our line of sight and the rotation axis, $\zeta$. For random samples of the angles $\alpha$ and $\zeta$ drawn from a sinusoidal distribution, the $W$ distribution exhibits a peak at $2\rho$. Finally therefore,
\begin{equation}
r_{\rm 90} = cP(W)^2/ 18 \pi.
\label{eqn:r90}    
\end{equation}

Complementing geometrical methods, the model of \cite{bcw91} enables emission height determination using aberration--retardation effects:
\begin{equation}
    r_{\rm bcw} = \frac{Pc\Delta\phi}{4 \cdot 360},
    \label{eqn:rbcw}
\end{equation}
where \(\Delta\phi\) is the phase shift (in degrees) between the profile midpoint and the inflexion of the position angle swing, $\phi_0$.
The determination of $\phi_0$ comes via the rotating vector model (RVM), originally proposed by \cite{rc69a} to describe the observed polarization angle (\(\Psi\)) as arising from the projection of a dipolar magnetic field as the pulsar's emission beam sweeps across an observer's line of sight. The characteristic polarization angle swing is given by
\begin{equation}
\Psi = \Psi_0 + \arctan \left( \frac{\sin \alpha \sin (\phi - \phi_0)}{\sin \zeta \cos \alpha - \cos \zeta \sin \alpha \cos (\phi - \phi_0)} \right),
\end{equation}
where $\phi$ is the rotational phase, and \(\phi_0\) is the phase at which the polarization angle \(\Psi\) passes through the fiducial plane. 

We first performed phase alignment of the subband profiles. Initially, the peak of the profile in subband 3 is set to phase 180 deg. Then, phase alignment and polarization angle alignment of the pulsar profiles across the subbands were obtained using cross-correlation techniques. The horizontal alignment ensures proper phase alignment across frequency bands, while the vertical alignment corrects for frequency-dependent Faraday rotation effects.

The RVM fitting procedure employed a two-stage optimization approach. Initially, a global fit was performed across all frequency bands using differential evolution to avoid local minima, followed by a local optimization using the L-BFGS-B algorithm. These algorithms were implemented through SciPy's optimization module \citep{vir20}. This hybrid approach ensured robust determination of the key geometric parameters $\alpha$, $\beta$, $\phi_0$ and $\psi_0$. Parameter uncertainties were estimated using the Hessian matrix. The emission height was derived from the phase shift ($\Delta \phi_0$) between $\phi_0$ as determined from the RVM fit and the center of the profile measured at 50\% of the peak intensity. The uncertainties in the emission heights was propagated from the uncertainties in $\Delta \phi_0$, which was a combination of the error from the RVM fitting and the profile center with an uncertainty of 1\degr \footnote{An uncertainty of 1\degr is a conservative estimate based on 1024-bin pulse profiles.}, such that $\sigma(\Delta \phi_0) = \sqrt{ \sigma^2(\phi_0) + 1 }$. 

\section{Results}
In this section we show figures for one pulsar, PSR~J0738--4042 as illustrative of the sample as a whole. We note that although this pulsar has undergone a slow profile change over time \citep{bkb+14,zgy+23,ljk+23}) this does not affect the results shown here. The figures for the other pulsars are contained in the supplementary material. 

\subsection{Profile widths}
\subsubsection{Fraction of Peak intensity}
\label{sec.intensity}
The profile width evolution for \object{PSR~J0738--4042} is shown in Fig.~\ref{fig:0738}. The profiles are displayed in the left panels, with different colors corresponding to the different subbands. The profile width versus frequency at different intensity percentages is shown in the middle panels. The fitted solutions, $A$, $\mu$ and $W_{x,0}$, are shown in the right panels. The middle panel shows some of the issues inherent in the fitting. The width evolution with frequency at the various percentage cuts depends critically on whether the leading component is included or not. When it is not included the frequency evolution is essentially flat, $A$ is zero and $\mu$ is undefined. This finding is repeated in many of the pulsars in our sample, particularly those with a large number of components.

For the 144 pulsars which have $\sigma_{\mu}< 2$\degr, there are 56 negative and 25 positive values of $\mu$ at $W_{10\%}$, and 71 negative and 33 positive values of $\mu$ at $W_{50\%}$. A two-dimensional contour plot of $\mu$ for all the pulsars in the sample as a function of intensity levels is shown in Fig.~\ref{fig:widthhistoweight}. The counts are weighted by $\sigma_{\mu}$. The figure indicates a tendency towards negative $\mu$ for $x < 50\%$ although approximately 30\% of the sample are at positive values of $\mu$. Above $x > 50\%$, the tendency is weaker, indicating that the method breaks down. This is particularly the case for multi-component profiles, where the high-intensity cut-off misses weak components at the profile edges. Our findings are in agreement with the frequency-dependent behaviors previously documented by \citet{cw14} and \citet{pkj+21}. This corroborates the notion that, while profile narrowing at higher frequencies remains prevalent, a significant proportion of pulsars manifest contrasting width evolution patterns.

In summary, although historically used and easily applied with tabulated values, a method based on the full width at half maximum ($W_{50}$), or more generally the full width at x maximum ($W_{x}$), is not robust to the choice of x, especially for $x>20\%$. 
This result is visible in Fig.~\ref{fig:widthhistoweight}, which shows that the apparent frequency mapping of the peaks becomes more dispersed as x increases. The apparent width evolution is therefore (in many cases) not dependent on any physics to do with diverging field-lines or the emission process.

\subsubsection{Component separation}
\label{sec.compsep}
\begin{figure}
    \centering
    \includegraphics[width=1\linewidth]{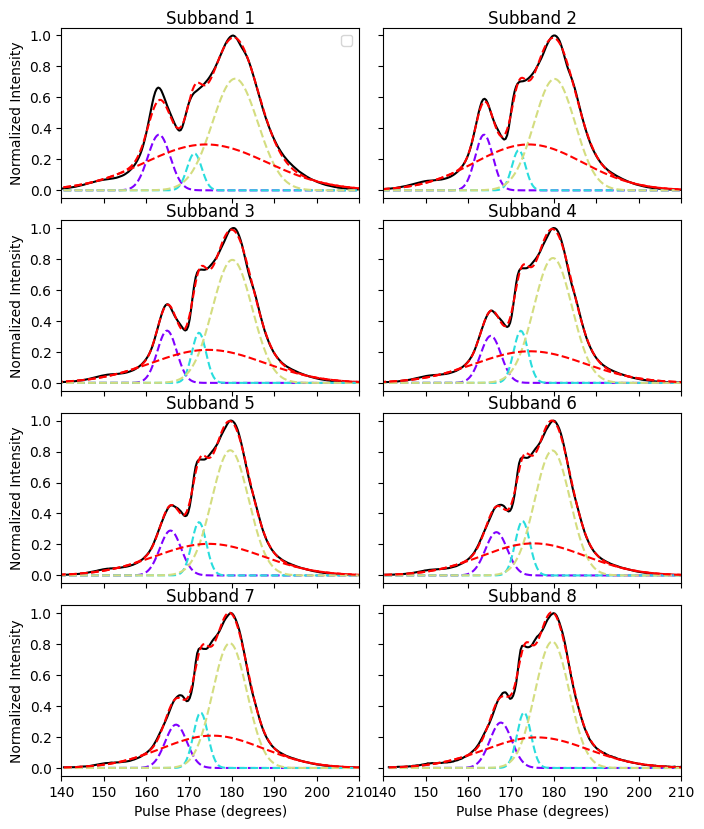}
    \caption{Gaussian component fits of PSR~J0738--4042 over 8 subbands, showing 4 individual components (dashed red, blue, purple, and yellow). The reconstructed profile is shown in dotted red over the profile (black).}
    \label{fig:fitgauss}
\end{figure}

\begin{figure}
    \centering
    \includegraphics[width=1\linewidth]{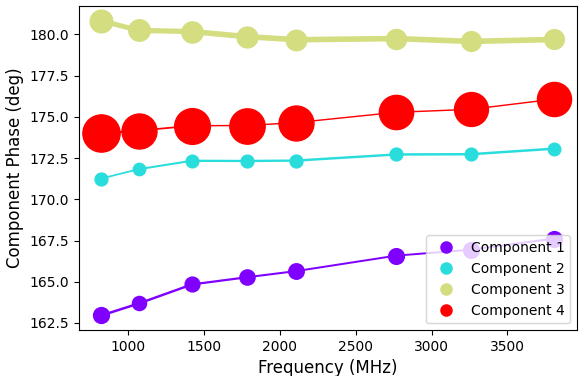}
    \caption{Fitting results of PSR~J0738--4042. The y-axis and x-axis represent the Gaussian component location over frequency. The size of the circles represent  the widths of the components.}
    \label{fig:gausscomp}
\end{figure}

\begin{figure}
    \centering
    \includegraphics[width=1\linewidth]{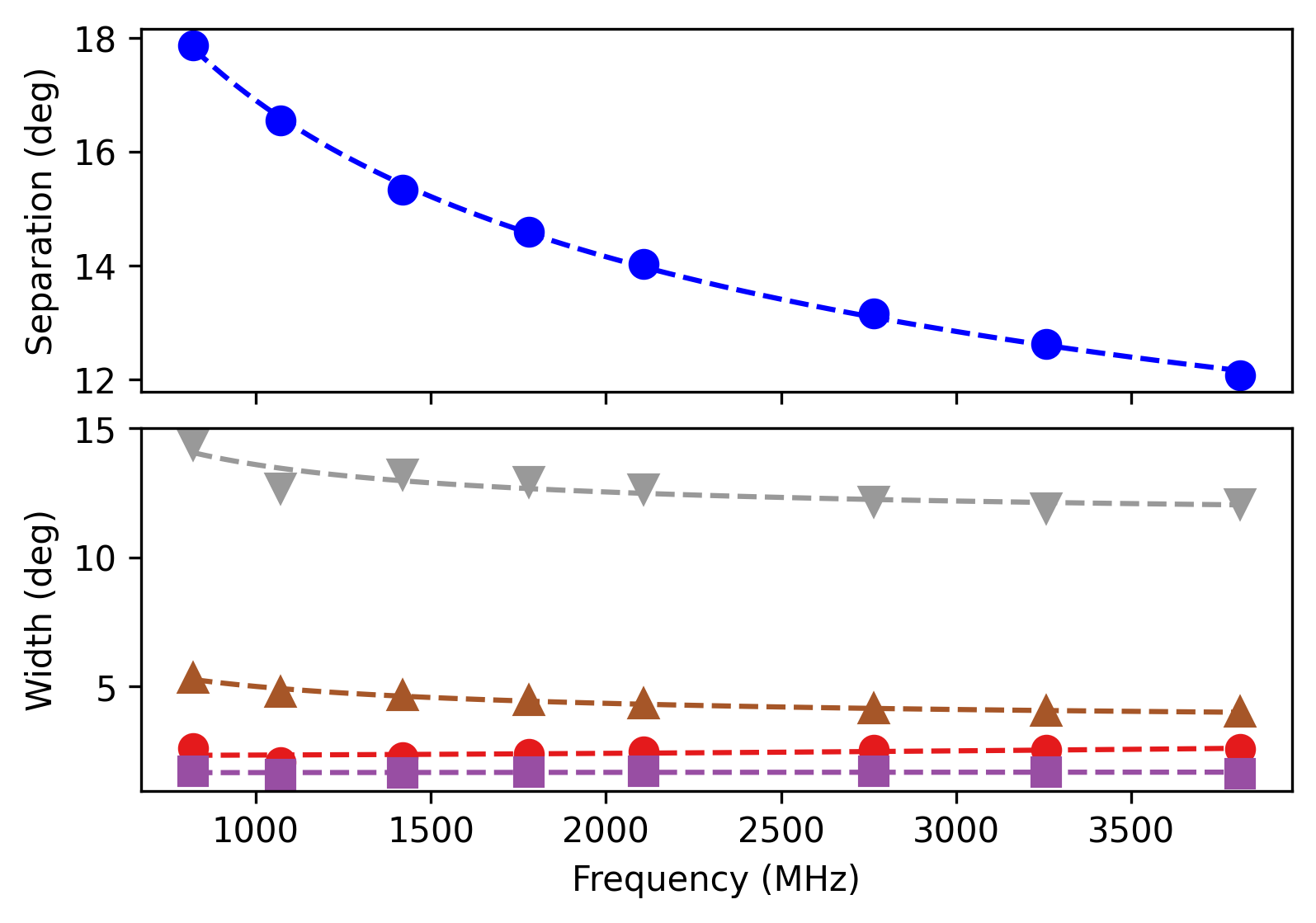}
    \caption{Frequency evolution of the maximum component separation ($top$) and the component width ($bottom$) of PSR~J0738--4042. The spectral lines are shown with best fit solutions of Eqn.~\ref{eq:thorsett}.   
    }
    \label{fig:gausscompthorsett}
\end{figure}

An example of the component fitting for PSR~J0738--4042 is shown in Figs.~\ref{fig:fitgauss} and \ref{fig:gausscomp}. The Gaussian reconstruction consistently traces the evolution of the component with frequency, with each component maintaining its identity across all subbands. The upper panel of Fig.~\ref{fig:gausscompthorsett} displays the maximum component separation ($W_{\rm comp.sep}$) as a function of frequency, along with the best fit to Eqn.~\ref{eq:thorsett}.

\begin{figure}
    \centering
    \includegraphics[width=1\linewidth]{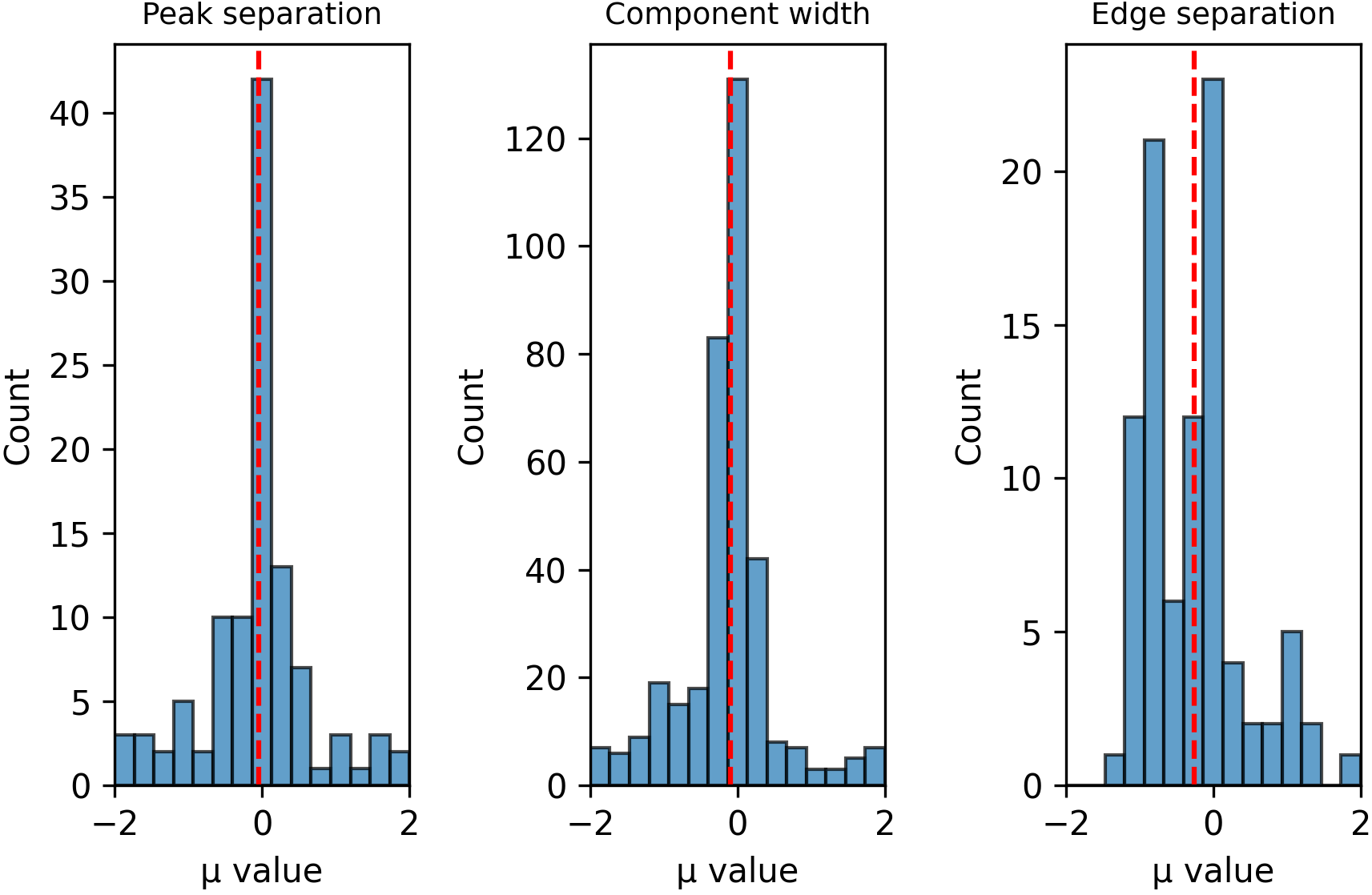}
    \caption{Histogram of $\mu_{\rm comp.sep}$ (left), $\mu_{\rm comp.width}$  (middle) and $\mu_{\rm comp.edge}$ derived from fitting the maximum peak-component separation, the individual component widths, and the maximum edge-component separation with Eqn 1. See text for detail.}
    \label{fig:gausscomphisto}
\end{figure}

The distribution of $\mu_{\rm comp.sep}$ across 128 pulsars is shown in Fig.~\ref{fig:gausscomphisto} (left panel). This distribution is approximately symmetric and centered near zero, with a median $\mu_{\rm comp.sep}$ value of $-0.04 \pm 0.48$. The uncertainty is the $\pm1\sigma$ percentile of the distribution. The clustering around zero suggests that for many pulsars, the separation between components remains relatively stable across frequency. There is a slight bias toward negative values, suggesting a weak tendency for components to converge with increasing frequency. 

\subsubsection{Component width}
\label{sec.compwidth}

To illustrate the frequency dependence of component widths, Fig.~\ref{fig:gausscompthorsett} (bottom panel) presents the widths of the four Gaussian components identified in PSR~J0738--4042 (Figs.~\ref{fig:fitgauss} and~\ref{fig:gausscomp}) as a function of frequency.
The distribution of $\mu_{\rm comp.width}$ across 128 pulsars is shown in Fig.~\ref{fig:gausscomphisto} (middle panel). The distribution is approximately centered near zero, with a median $\mu_{\rm comp.width}$ value of $-0.10 \pm 0.39$. This value is more negative than that of the component separation mean ($\mu_{\rm comp.sep} = -0.04 \pm 0.48$) suggesting a stronger tendency for individual components to narrow with increasing frequency. There is a high concentration of components with $\mu_{\rm comp.width}$ values near zero. 

To confirm the distinction between the evolution of the component width (4.1.3) and  the component peak separation (4.1.2), we derive an additional width parameter, $W_{\rm comp.edge}$, which is a slight modification to $W_{\rm comp.sep}$. The position of the edge of an individual Gaussian component is measured at 20\% of the peak amplitude of the full profile. For each subband, the largest separation between those edges, i.e. the leftmost and the rightmost components edges of profile, defines the width $W_{\rm comp.edge}$. Then, for each pulsar, $W_{\rm comp.edge}$ of the subbands are fitted using Eqn. 1, following the third method in Section 3.1. 

The results are shown in Fig.~\ref{fig:gausscomphisto} (right), which has a median $\mu_{\rm comp.edge}$ of $-0.26 \pm 0.53$. The distribution, which has two peaks at $\sim$0 and $\sim -1$, indicates an even stronger tendency to narrow with increasing observing frequency. 
In method 1, $W_{\rm comp.sep}$ uses the peak position of the outermost Gaussian components, resulting in minimal to no frequency dependency (left). However, for $W_{\rm comp.edge}$, where the boundary is changed to the 20\%-peak edge position, the distribution considerably shows a stronger frequency dependency in the width evolution (right). These results suggest that the frequency evolution of the profile width is dominated by the frequency evolution of the width of the individual Gaussian components.

\subsection{Emission heights}

\begin{figure}
    \centering
    \includegraphics[width=1\linewidth]{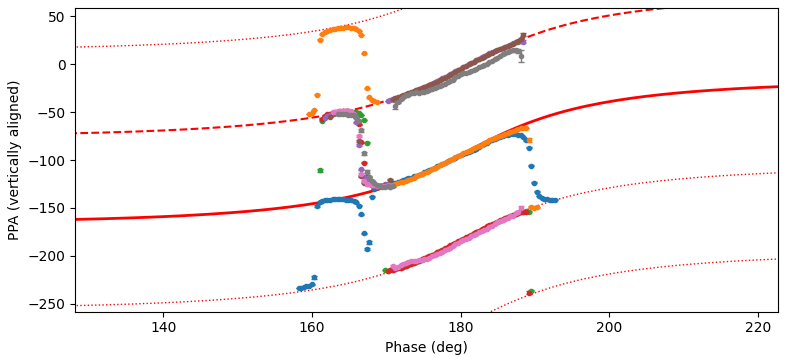}
    \includegraphics[width=1\linewidth]{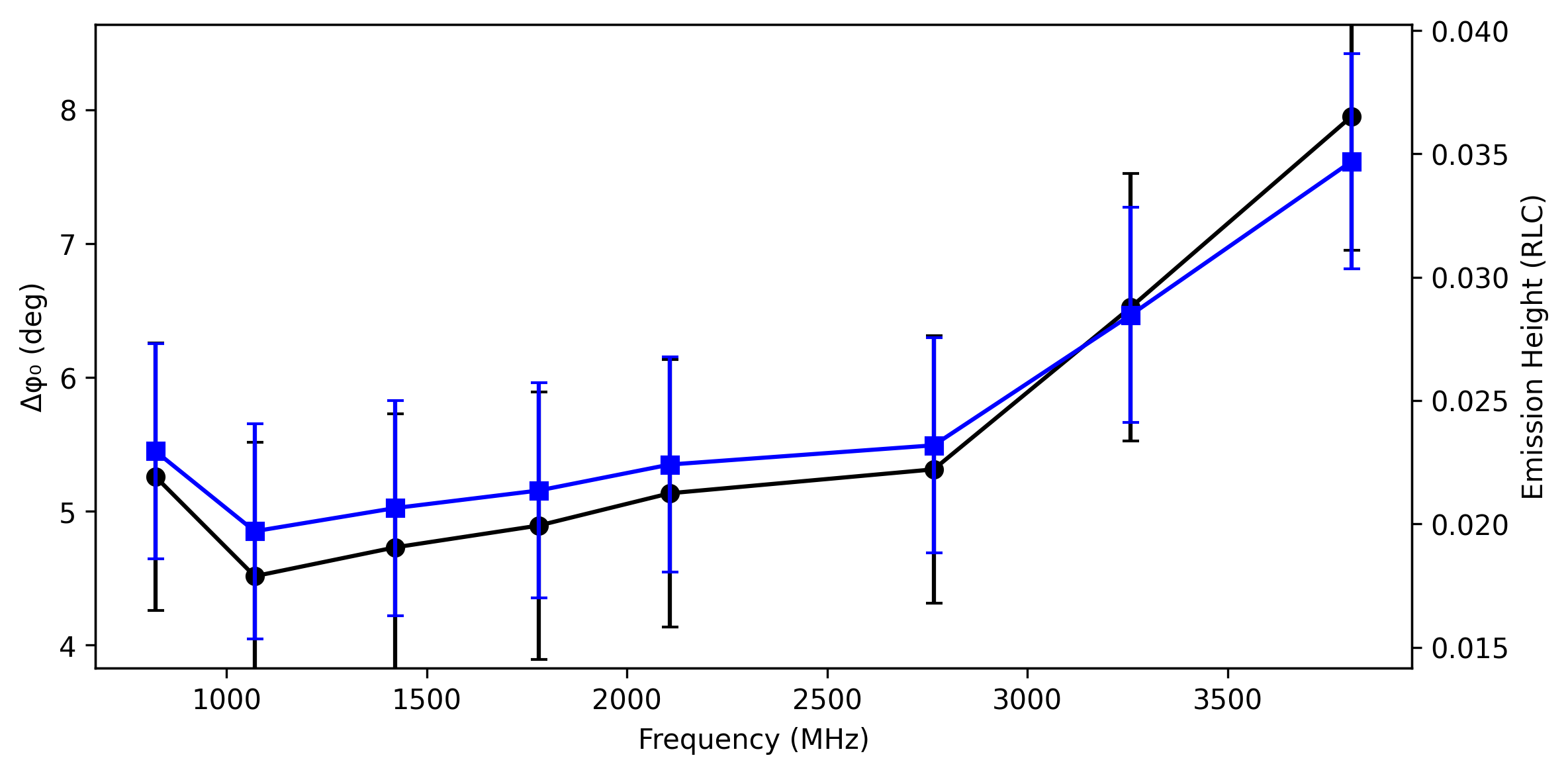}
    \caption{PPAs of PSR~J0738–-4042 across eight frequency bands. The PPAs are phase-aligned, and position angle swings are overlaid with best-fit RVM curves.     
    The bottom panel shows $\Delta \phi_0$ (black) and the emission height (blue) against frequency, both appear to increase with frequency.    }
    
    \label{fig:0738bcw}
\end{figure}

Fig.~\ref{fig:0738bcw} presents the aligned polarization position angle (PPA) sweeps for PSR~J0738–4042 across the observed subbands, demonstrating the RVM fits. The evolution of emission heights with frequency for this pulsar is measured as a fraction of the light cylinder radius ($R_{\rm LC}$).

Following the criteria by \cite{jmk+24}, pulsars are classified into three groups based on their $\Delta \phi_0$: Group -1 for $\Delta \phi_0 < -1^{\circ}$, Group 0 for $-1^{\circ} \leq \Delta \phi_0 \leq 1^{\circ}$, and Group 1 for $\Delta \phi_0 > 1^{\circ}$. Based on this classification, our dataset contains 83 pulsars in Group 1, 51 pulsars in Group -1, and 23 pulsars in Group 0. This distribution suggests that the majority of pulsars exhibit a significant positive phase shift (68\% for Group 0 and 1), while approximately a third display strong negative shifts (32\%). 

The same results are divided into positive and negative trend with frequency, which are 82 (52\%) and 75 (48\%), respectively. 

\begin{figure}
    \centering
    \includegraphics[width=1\linewidth]{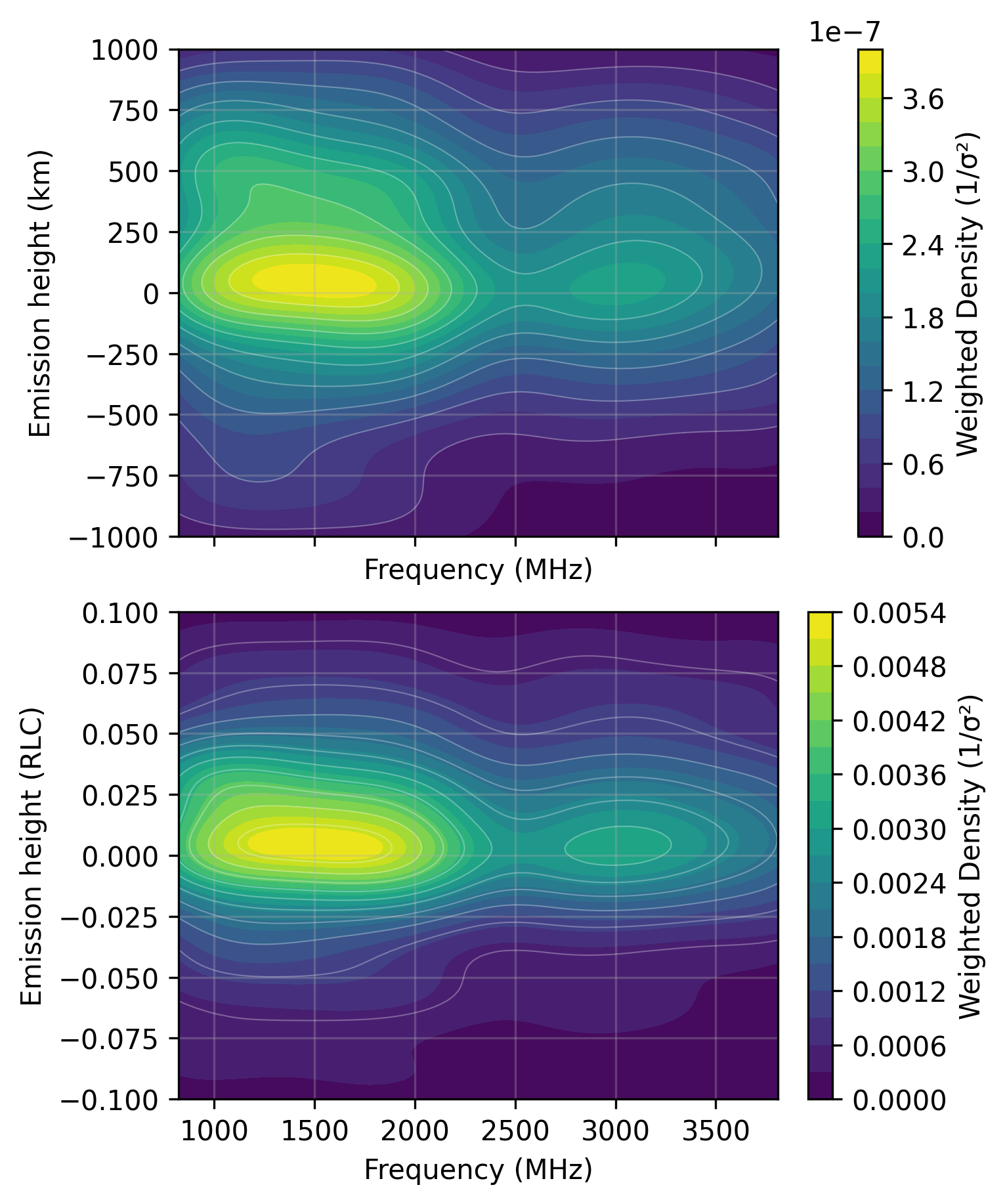}
    \caption{Emission height $r_{\rm bcw}$ vs. $\nu$ (1000-3500 MHz)
   showing weighted density distributions. The panels display emission heights in km (top) and light cylinder radii $R_{\rm LC}$ (bottom). The color gradient represents probability density weighted by $1/\sigma^2$ of individual measurements, highlighting regions with higher measurement confidence.    
    }
    \label{fig:combinedheight}
\end{figure}

Fig.~\ref{fig:combinedheight} shows how $r_{\rm bcw}$ changes with observing frequency. The top panel plots $r_{\rm bcw}$  in km, while the bottom panel is in units of $R_{\rm LC}$. Each panel uses a color map to represent the density of data points and is weighted by $1/\sigma^2$, where $\sigma$ is the measurement error. Most pulsars cluster around heights of 100 km or about 0.01 $R_{\rm LC}$, particularly in the 1200–1800 MHz range. The emission pattern generally shows a symmetric distribution around the 100 km height, with contour lines indicating the emission density constant with frequency.

\subsection{Relationship with $\dot{E}$}

\begin{figure}
    \centering
    \includegraphics[width=1.05\linewidth]{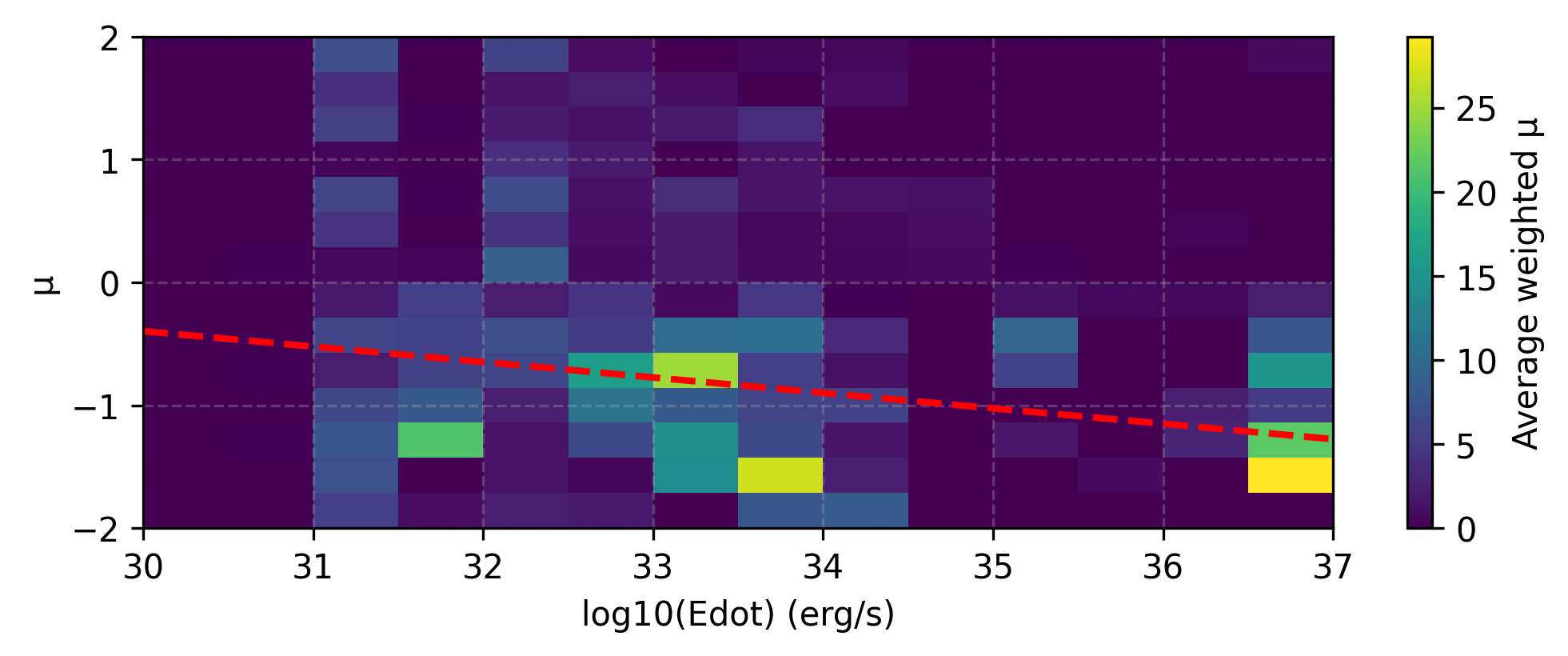}
    \caption{Error-weighted distribution of $\mu$ as a function of $\log\dot{E}$. The color scale represents normalized weighted density, with brighter colors indicating higher density regions. A weighted linear regression demonstrates a moderate negative correlation between profile shape and spin-down energy loss rate. Only measurements with $-2 < \mu < 2$ are included in this analysis.  }
    \label{fig:muedot}
\end{figure}

We now examine how pulsar spin-down luminosity, \(\dot{E}\), relates to various emission properties. Fig.~\ref{fig:muedot} shows the two‐dimensional density plot of the profile width evolution parameter, \(\mu\), against \(\log(\dot{E})\). A weighted linear regression yields $\mu = -0.16 \log(\dot{E}) + 4.8 \quad($coefficient of determination, $R^2 = 0.5)$, indicating that pulsars with higher \(\dot{E}\) generally exhibit stronger profile narrowing with increasing frequency. 
To test the robustness of this trend, we computed the Spearman rank correlation between $\mu$ and $\log_{10}(\dot{E})$, obtaining $\rho = -0.37, p = 0.22.$ While the negative $\rho$ suggests a weak‐to‐moderate inverse monotonic relationship, the high $p$-value implies that this correlation is not statistically significant. 
The data are concentrated around \(\mu \lesssim 0\) and \(\dot{E}\) in the range \(10^{33}\) to \(10^{34}\) erg/s.

\begin{figure}
    \centering
    \includegraphics[width=1.05\linewidth]{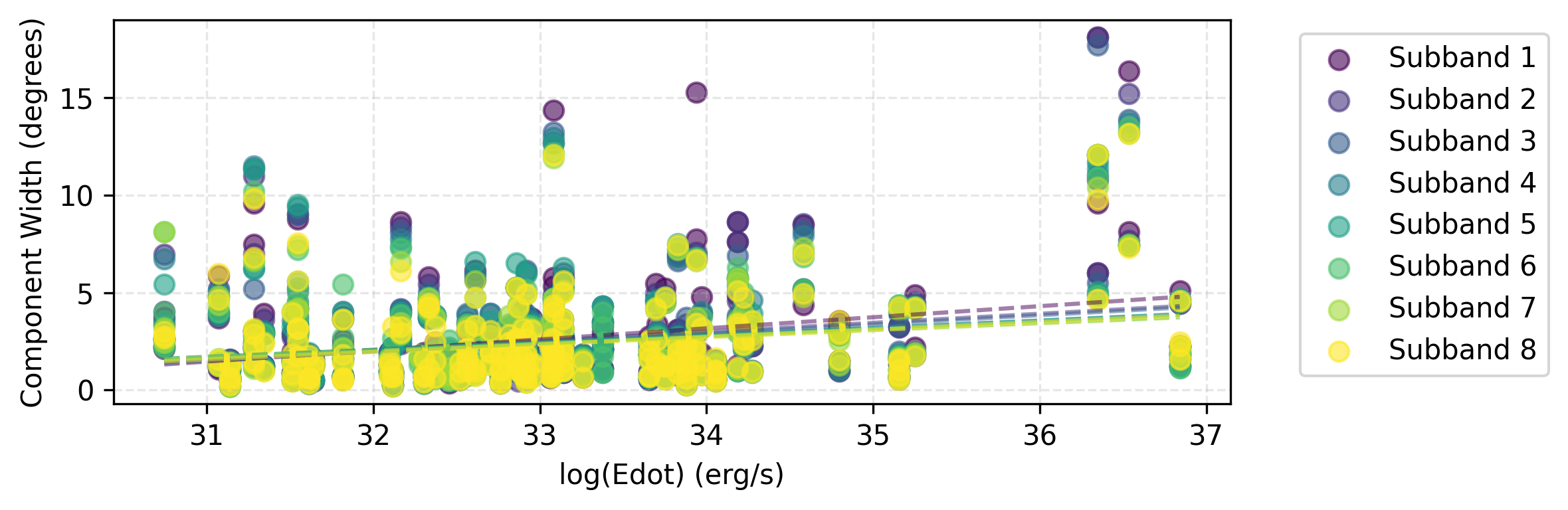}
    \caption{Relationship between the component width (deg) and $\log(\dot{E})$ (erg/s) across eight frequency subbands. Each subband is represented by a different color according to the viridis colormap, with corresponding linear regression fits shown as dashed lines. The data reveal a positive correlation,  suggesting that pulsars with higher $\dot{E}$ tend to exhibit wider emission components. The scatter in component widths is more pronounced at higher $\dot{E}$ values, particularly above $10^{36}$ erg/s.}
    \label{fig:mumethod3}
\end{figure}

To further dissect this behavior, we extended our analysis to include the evolution of Gaussian component parameters. In Fig.~\ref{fig:mumethod3}, the component width evolution parameter, \(\mu_{\text{comp.width}}\), shows a weak positive correlation with \(\log(\dot{E})\). 
The Spearman rank correlation test results in $\rho$ = 0.1511 and $p$-value = $1.31e^{-10}$ indicating a statistically significant but quantitatively small positive monotonic relationship between component width and $\log\dot{E}$.
This suggests that pulsars with higher $\dot{E}$ tend to have marginally wider individual emission components. Although, the regression slopes for different frequency subbands are similar, systematic offsets are observed: lower-frequency subbands generally yield wider component widths compared to higher-frequency subbands at similar \(\dot{E}\) values.
In contrast, our analysis of the component separation evolution parameter, \(\mu_{\text{comp.sep}}\), reveals no significant correlation with \(\dot{E}\). This indicates that the spacing between Gaussian components remains largely invariant with respect to the $\dot{E}$, and thus contributes less to the overall frequency evolution of the profile.

In addition, our analysis of $r_{\rm bcw}$ reveals that the majority of pulsars cluster at moderate heights (200--600~km) over a broad range of \(\dot{E}\) (\(10^{32}\)–\(10^{35}\) erg/s). In contrast to the trends seen in profile and component width evolution, $r_{\rm bcw}$ shows no significant dependence on spin-down luminosity. This finding supports earlier suggestions (e.g., \citealt{rwj+17,wj08}) that high-\(\dot{E}\) pulsars might develop wider profiles not because of a uniformly higher emission altitude, but due to emission occurring over an extended range of altitudes.

\begin{figure}
    \centering
    \includegraphics[width=1\linewidth]{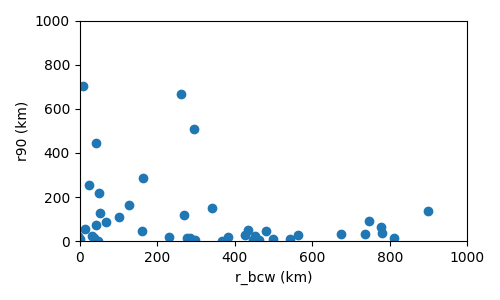}
    \caption{Scatter plot comparing emission altitudes derived from two different methods: $r_{\rm 90}$ in km versus $r_{\rm bcw}$ in km, showing a distribution of values up to 1000 km for both parameters."}
    \label{fig:rbcw_r90}
\end{figure}

\subsection{Data}
Table 1 in the Appendix lists the frequency evolution index ($\mu$) derived from three methods: total profile width at 20\% intensity ($\mu_{20}$), component separation ($\mu_{\rm comp.sep}$), and individual component width ($\mu_{\rm comp.width}$), each with uncertainties. 
Table 3 presents the group classification by $\Delta \phi_0$ the slope of height versus frequency and the correlation between emission height and observing frequency.
These data show that many pulsars deviate from the expected trend of decreasing height with frequency, highlighting the complexity of the emission geometry.

\section{Discussion}
\subsection{Profile widths}

Our analysis reveals a complex relationship between profile width and observing frequency that cannot be fully captured by traditional width metrics alone (width $\propto \nu^a$). The majority of pulsars in our sample exhibit a decrease in profile width with increasing frequency, consistent with older studies (e.g., Thorsett 1991; Mitra \& Rankin 2002), while a significant subset ($\sim$31\%) display an increase in profile width with frequency, consistent with more modern studies (\citealt{cw14,pkj+23}).

We caution that traditional width measurements, such as W$_{10}$ or W$_{50}$, fail to reveal the underlying frequency evolution of pulsar profiles, as demonstrated in Fig. 2. This figure supports the Thorsett relationship and RFM to some extent, but also clearly shows that one might reach contradictory conclusions depending on which intensity level is chosen. These traditional width metrics were established in early studies (\citealt{tho91}; \citealt{kwj+94}) when pulsar surveys were biased toward low frequencies (typically 400~MHz) and dominated by pulsars with prominent central components having steep spectral indices. However, our current study at higher frequencies (0.7-4.0 GHz) display profiles with outrider components which often dominate the profile, necessitating a different approach to the analysis.

A more insightful perspective emerges from analyzing the frequency evolution of the properties of the Gaussian components. We have shown that the separation of the outer components varies only weakly with frequency, whereas the widths of individual components do indeed narrow with increasing frequency. This therefore suggests that the increase in profile width with decreasing frequency is a propagation effect and does not reflect the divergence of the magnetic field lines or this may result from emission physics (e.g. narrower 1/gamma pencil beamlets).

These findings align with the fan beam model, which predicts minimal frequency evolution along the diverging radial fan beams of emission. In this model, refraction in wave propagation (particularly of the O-mode, \citealt{cr79,bar86}) manifests as widening of individual components. Rather than simply equating low frequency with high emission altitude, our analysis suggests that the relationship between frequency and emission properties is more nuanced and depends on the specific emission geometry and propagation effects. Given these complexities, we recommend that future studies focus on Gaussian component properties rather than width measurements such as W$_{10\%}$ that are heavily influenced by spectral characteristics of individual components. Component separation and width measurements provide more physically meaningful parameters that better reflect the underlying emission processes in pulsar magnetospheres.

\subsection{Emission height}
The results of Section 4 demonstrate an apparent dichotomy in emission height behavior (both $r_{\rm bcw}$ and $r_{90}$) across frequency bands. 
Determining $r_{\rm bcw}$ shows that, while 75 pulsars (48\%) exhibit a decreasing trend with frequency, 82 pulsars (52\%) show an apparent increase in emission height with frequency. 
Fig.~\ref{fig:rbcw_r90} shows $r_{\rm bcw}$ versus $r_{\rm 90}$ for our dataset. There is little correlation between the two values, with a Spearman rank-order correlation coefficient of $-0.010$. This is consistent with the conclusion reached by \citet{ml04}, \citet{wj08} and \citet{dkl+19}.

Therefore, when interpreting these results, we must acknowledge the inherent limitations in both methodologies employed to calculate emission heights. First, $r_{\rm bcw}$ depends critically on the accurate identification of the centre of the profile, which in our analysis is taken as the midpoint between the leading and trailing edges at 50\% of the maximum intensity. This approach introduces potential biases, particularly in asymmetric profiles and those with frequency-dependent component evolution. In addition the profile center measurement can be skewed due to incomplete beam illumination. Secondly, the geometric height calculations ($r_{90}$) rely on two critical assumptions: they depend on knowing the pulsar geometry ($\alpha$ and $\beta$), which remains poorly constrained for most pulsars (e.g., \citealt{ew01}, \citealt{jkk+23}) and they assume the beam is uniformly filled from edge to edge, which may not reflect actual emission patterns (e.g. \citealt{lm88}).

We note that \cite{dp15} compared fan beam models with the conventional conal models. In the fan beam picture, broadband and coherent emission from secondary relativistic particles forms radially extended sub-beams (\citealt{Wang2014,Saha2017,Huang2020}). When only one or a few flux tubes are active, the fan beam becomes patchy. \cite{dp15} concluded that the fan beam model gave a better fit to the data.

\subsection{Edot}
The profiles of high $\dot{E}$ pulsars are different to those of low $\dot{E}$ pulsars. Their components start wider at low frequencies but narrow more quickly as frequency increases. This pattern arises because high $\dot{E}$ pulsars tend to have smaller magnetospheres. With a smaller magnetosphere, the emission streams lie closer together. This proximity results in a stronger frequency dependence in the stream widths compared to normal pulsars. Other differences include that the emission from high $\dot{E}$ pulsars may come from higher altitudes \citep{jw06}, over a large range of emission heights \citep{kj07} and possibly even from outside the conventional polar cap \citep{ww09,rwj15}. This could all contribute to profiles that appear wider at low frequencies (as the fan beam streams along the field lines) but experience a sharper contraction with increasing frequency.

This behavior inspires the fan-beam model in the next section. Each emission stream changes its width more rapidly with frequency in high $\dot{E}$ pulsars. Their compact magnetospheres produce streams that are more sensitive to frequency-dependent effects. As a result, the individual components narrow faster as frequency increases.

\begin{figure*}   
    \centering
    \includegraphics[width=0.9\linewidth]{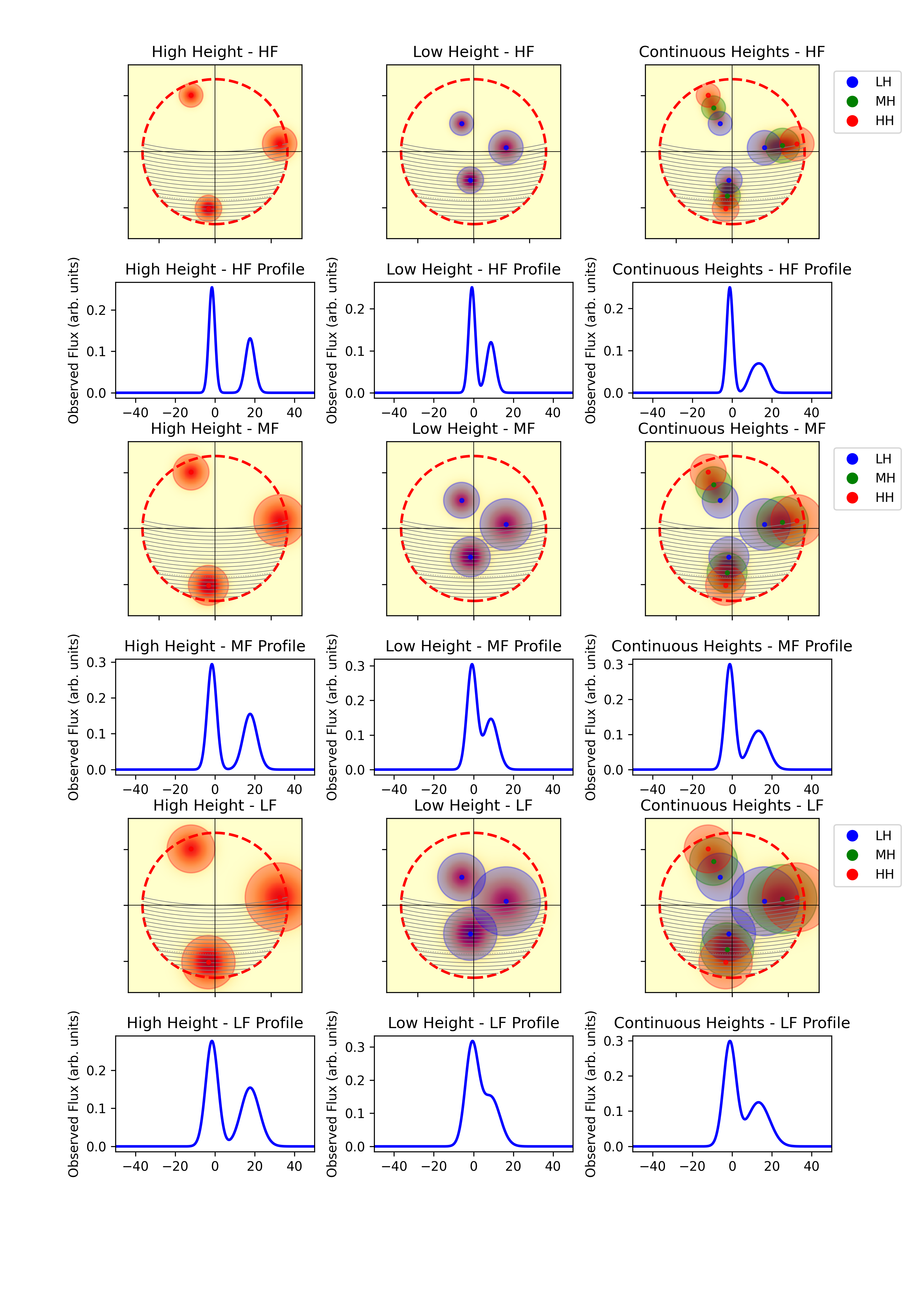}
    \caption{Cartoon of a fan beam model. Beam maps (top row in each frequency band) and profiles (bottom row in each frequency band) for three height configurations arranged by column: high height (left), low height (center), and multiple heights (right). From top to bottom, each pair of rows corresponds to a different frequency band: high frequency (HF), medium frequency (MF), and low frequency (LF). The dashed red circle indicates the emission boundary, and the color scale denotes relative flux intensity. The profile illustrates how amplitude varies with horizontal displacement. See text for details.}
    \label{fig:beammodel}
\end{figure*}

\subsection{A Fan Beam Model}
Based on the results outlined in Sections 4.1 and 4.2, we propose a simple fan-beam model to explain the data.

Fig.~\ref{fig:beammodel} illustrates the fan-beam model under different assumptions of emission height and observing frequency. Three distinct columns represent high emission height (left), low emission height (middle), and a summation across a continuous range of emission heights (right). In each panel, the line of sight (LOS) is confined to the lower half of the beam, as indicated by the faint dotted lines filling the bottom region of the circular emission zone. This is because the fan beams stream towards the observer in all LOS-parallel directions. The rows represent different observing frequencies, with the top row corresponding to higher frequencies (HF) and the lower rows corresponding to progressively lower frequencies (MF, LF).

A key outcome from this model is that individual emission streams, which are presumed to follow the open magnetic field lines, are broader at low frequency and narrower at high frequency. In contrast to simple geometric models with diverging field lines, the variation in beam width here arises primarily from intrinsic frequency-dependent effects (e.g., refraction or plasma frequency–dependent broadening; \cite{mck97}, \cite{pk22}) rather than merely from the changing spread of field lines with emission height. Thus, the fan-beam interpretation allows a more nuanced view: the overall beam shape changes modestly with height, but each emission “blob” widens significantly at lower frequencies.

The summation of emission over a continuous range of heights (right column) reproduces, at least qualitatively, the beam patterns observed in more complex pulsar systems, such as the precessing pulsar PSR~J1906+0746 (\citealt{dkl+19}), where a streak of emission from the magnetic axis is observed. The beam map retains a characteristic asymmetry and frequency evolution, implying that broadband emission originating over a wide range of altitudes can still manifest clear frequency-dependent beam widths.

An important consequence from the data analysis in Section 4.1 is that the profile widths evolve because of changes in the component widths rather than changes in the centroids of the components. As can be seen in Fig.~\ref{fig:beammodel}, while the overall beam broadens at low frequency, the centroids of the individual emission streams do not shift substantially. In addition, we found that emission heights have virtually no frequency dependence. Within the framework of the model, this arises because the observer sees emission from multiple altitudes at the same time. 

Finally, the standard RVM remains largely unaffected by the details of emission height in this fan-beam picture. All field lines, regardless of altitude, intersect the observer’s line of sight at effectively the same magnetic-polar angle. Consequently, the basic RVM sweep is preserved even as the emission streams widen or narrow with frequency, implying that polarization angle constraints on the magnetic geometry remain robust. Overall, the results emphasize that broadband fan-beam emission, spanning multiple heights, can naturally produce wider observed profiles at low frequencies without significantly altering the geometric center or the position-angle swing.

\section{Conclusions}
We have conducted an investigation into the frequency evolution of pulsar emission geometry using data from the Murriyang (Parkes) telescope. By examining both profile width and emission altitude across a sample of more than 100 pulsars, our results reveal a more intricate picture than that suggested by the simple RFM model. While many pulsars show the classical trend of narrower profiles and lower emission heights with increasing frequency, a significant subset exhibits the opposite behavior i.e. broadening profiles or higher emission altitudes at higher frequencies.

In particular, we find that measuring total profile width (e.g., $W_{10}$ or $W_{50}$) can obscure the underlying component-level evolution. A more robust view emerges by fitting and tracking individual Gaussian components. Most pulsars do not show large shifts in the centroids of these components; instead, the component widths themselves evolve strongly with frequency. This suggests that the radiation is not simply emitted from a single, well-defined conal boundary but may instead arise from fan-like streams whose intrinsic width and intensity vary with observing frequency.

Our measured emission heights, based on aberration–retardation methods, reveal no single, universal pattern. We point out that the assumptions inherent in the method are largely obscuring the true picture. Moreover, we find no clear correlation between emission altitude and spin-down luminosity ($\dot{E}$), even though pulsars with larger $\dot{E}$ show more pronounced frequency evolution in their profile widths. Together, these observations imply that pulsar radio emission arises from a broad, complex region within the magnetosphere, rather than from a thin, discrete layer.

It is recommended that future research explore the potential of simultaneous multi-frequency single-pulse data (e.g.\citealt{okj19, jmk+24,mbm24}) in order to ascertain further evidence of the beam geometry.

\section*{Data availability}
The pulsar data is available to download via the CSIRO Data Access Portal (\href{https://data.csiro.au/}{https://data.csiro.au/}). Figures for individual pulsars, similar to those described for PSR~J0738--4042 in Section 4, can be found at https://doi.org/10.5281/zenodo.16308954.

\begin{acknowledgements}
This work is supported by the Fundamental Fund of Thailand Science Research and Innovation (TSRI) through the National Astronomical Research Institute of Thailand (Public Organization) (FFB680072/0269).
Murriyang, the Parkes radio telescope, is part of the Australia Telescope National Facility (https://ror.org/05qajvd42) which is funded by the Australian Government for operation as a National Facility managed by CSIRO. We acknowledge the Wiradjuri people as the traditional owners of the Observatory site. MEL is supported by an Australian Research Council (ARC) Discovery Early Career Research Award DE250100508. Work at NRL is supported by NASA.
\end{acknowledgements}

\bibliographystyle{aa}
\bibliography{psrrefs}

\begin{appendix}
\section{Results from Section 4.1.}
\begin{table*}[!b]
\caption{Resulting frequency evolution parameters for profile widths.}
\centering
\begin{tabular}{l|rr|rr|rr|l|rr|rr|rr}
\hline\hline
Pulsar Name & $\mu_{20}$ & $\sigma_{20}$ & $\mu_{\rm sep}$ & $\sigma_{\rm sep}$ & $\mu_{\rm width}$ & $\sigma_{\rm width}$ & 
Pulsar Name & $\mu_{20}$ & $\sigma_{20}$ & $\mu_{\rm sep}$ & $\sigma_{\rm sep}$ & $\mu_{\rm width}$ & $\sigma_{\rm width}$ \\
\hline
J0034$-$0721 & $-1.0$ & $0.4$ & $-2.2$ & $0.4$ & $0.0$ & $4.6$ & J1136$-$5525 & $-1.8$ & $4.4$ & $0.7$ & $1.2$ & $-0.1$ & $1.1$ \\
J0134$-$2937 & $-1.4$ & $0.2$ & $-0.4$ & $0.3$ & $-5.0$ & $1.4$ & J1146$-$6030 & $-1.1$ & $0.9$ & $-5.0$ & $8.1$ & $5.0$ & $19.0$ \\
J0151$-$0635 & $-3.3$ & $3.3$ & $-0.3$ & $0.5$ & $0.1$ & $7.8$ & J1157$-$6224 & $0.7$ & $0.5$ & $2.1$ & $1.6$ & $0.2$ & $1.1$ \\
J0152$-$1637 & $-0.2$ & $0.1$ & $-$ & $-$ & $-$ & $-$ & J1210$-$5559 & $5.0$ & $1.9$ & $0.0$ & $2.5$ & $0.1$ & $2.3$ \\
J0206$-$4028 & $5.0$ & $10.2$ & $-$ & $-$ & $-$ & $-$ & J1224$-$6407 & $0.0$ & $4.2$ & $-$ & $-$ & $-$ & $-$ \\
J0255$-$5304 & $1.8$ & $0.3$ & $0.6$ & $0.8$ & $-0.1$ & $1.4$ & J1225$-$6408 & $0.2$ & $0.8$ & $0.2$ & $0.8$ & $3.2$ & $2.9$ \\
J0304$+$1932 & $-1.4$ & $0.3$ & $-0.9$ & $0.2$ & $5.0$ & $6.3$ & J1243$-$6423 & $1.7$ & $6.5$ & $-0.1$ & $0.9$ & $5.0$ & $2.7$ \\
J0401$-$7608 & $-0.1$ & $0.4$ & $-0.1$ & $0.3$ & $-1.0$ & $0.6$ & J1253$-$5820 & $-1.2$ & $0.3$ & $0.6$ & $4.5$ & $-1.0$ & $0.4$ \\
J0448$-$2749 & $0.0$ & $3.9$ & $-2.1$ & $2.3$ & $0.0$ & $8.9$ & J1319$-$6056 & $5.0$ & $10.7$ & $3.7$ & $14.9$ & $-$ & $-$ \\
J0452$-$1759 & $0.1$ & $0.3$ & $0.1$ & $1.9$ & $-$ & $-$ & J1320$-$5359 & $1.8$ & $2.2$ & $1.1$ & $6.1$ & $0.0$ & $12.6$ \\
J0525$+$1115 & $1.9$ & $10.1$ & $0.0$ & $15.2$ & $0.0$ & $45.2$ & J1326$-$6408 & $-0.1$ & $3.0$ & $0.0$ & $71.4$ & $0.0$ & $14.6$ \\
J0536$-$7543 & $-0.1$ & $0.4$ & $-0.1$ & $2.0$ & $5.0$ & $1.9$ & J1326$-$6700 & $-0.7$ & $0.2$ & $-0.2$ & $0.1$ & $-0.3$ & $0.8$ \\
J0543$+$2329 & $0.1$ & $1.2$ & $0.0$ & $6.3$ & $-0.1$ & $1.0$ & J1327$-$6301 & $3.6$ & $1.9$ & $-0.2$ & $1.2$ & $5.0$ & $6.3$ \\
J0601$-$0527 & $-0.2$ & $0.4$ & $1.9$ & $1.7$ & $-$ & $-$ & J1328$-$4357 & $-0.8$ & $0.2$ & $-0.9$ & $1.3$ & $-1.8$ & $0.4$ \\
J0614$+$2229 & $0.0$ & $19.9$ & $0.0$ & $43.7$ & $5.0$ & $5.9$ & J1401$-$6357 & $0.0$ & $1.8$ & $0.0$ & $10.0$ & $0.1$ & $10.0$ \\
J0624$-$0424 & $5.0$ & $6.0$ & $-0.3$ & $3.9$ & $2.1$ & $2.6$ & J1424$-$5822 & $-0.5$ & $0.5$ & $-0.7$ & $3.9$ & $-1.1$ & $2.2$ \\
J0630$-$2834 & $-0.2$ & $0.7$ & $0.0$ & $3.0$ & $-0.1$ & $0.8$ & J1428$-$5530 & $-0.3$ & $0.1$ & $-0.4$ & $0.3$ & $-0.2$ & $0.9$ \\
J0631$+$1036 & $-0.6$ & $0.8$ & $-$ & $-$ & $-$ & $-$ & J1430$-$6623 & $-0.2$ & $1.9$ & $-$ & $-$ & $-$ & $-$ \\
J0659$+$1414 & $-5.0$ & $1.0$ & $0.4$ & $4.2$ & $-0.1$ & $1.5$ & J1453$-$6413 & $0.3$ & $1.1$ & $0.4$ & $2.6$ & $-0.1$ & $3.5$ \\
J0729$-$1448 & $0.2$ & $2.7$ & $0.0$ & $7.0$ & $0.1$ & $4.3$ & J1456$-$6843 & $-0.2$ & $0.4$ & $-0.1$ & $1.7$ & $-5.0$ & $3.2$ \\
J0729$-$1836 & $-0.1$ & $0.5$ & $-0.1$ & $0.5$ & $-0.2$ & $1.3$ & J1522$-$5829 & $-1.2$ & $0.5$ & $-0.1$ & $7.3$ & $-2.8$ & $0.5$ \\
J0738$-$4042 & $-1.3$ & $0.2$ & $-0.4$ & $0.1$ & $-2.5$ & $1.5$ & J1530$-$5327 & $-0.8$ & $2.4$ & $-0.1$ & $2.0$ & $-0.1$ & $1.9$ \\
J0742$-$2822 & $0.1$ & $1.9$ & $-0.1$ & $0.2$ & $0.0$ & $2.7$ & J1534$-$5334 & $-0.3$ & $0.7$ & $-1.7$ & $1.0$ & $-0.4$ & $0.5$ \\
J0745$-$5353 & $-5.0$ & $3.2$ & $0.3$ & $1.1$ & $-5.0$ & $1.8$ & J1535$-$4114 & $-0.3$ & $0.4$ & $-0.6$ & $0.3$ & $-0.1$ & $3.0$ \\
J0758$-$1528 & $-1.1$ & $0.4$ & $-0.4$ & $0.3$ & $-0.1$ & $1.7$ & J1536$-$5433 & $1.2$ & $0.7$ & $-$ & $-$ & $-$ & $-$ \\
J0809$-$4753 & $0.1$ & $2.3$ & $0.2$ & $3.0$ & $0.1$ & $2.1$ & J1544$-$5308 & $0.0$ & $12.4$ & $-0.2$ & $1.4$ & $0.0$ & $3.8$ \\
J0820$-$1350 & $0.0$ & $3.0$ & $-0.1$ & $10.0$ & $2.2$ & $2.4$ & J1555$-$3134 & $0.0$ & $2.5$ & $0.0$ & $12.4$ & $0.0$ & $14.6$ \\
J0835$-$4510 & $1.8$ & $1.0$ & $0.1$ & $2.6$ & $-5.0$ & $8.6$ & J1557$-$4258 & $0.7$ & $1.3$ & $0.4$ & $1.0$ & $0.0$ & $6.3$ \\
J0837$+$0610 & $2.1$ & $0.5$ & $2.1$ & $0.6$ & $-$ & $-$ & J1559$-$4438 & $0.5$ & $0.7$ & $0.1$ & $1.7$ & $-$ & $-$ \\
J0837$-$4135 & $2.2$ & $0.5$ & $5.0$ & $3.5$ & $0.6$ & $0.5$ & J1600$-$5751 & $0.7$ & $1.8$ & $0.2$ & $2.8$ & $0.0$ & $7.7$ \\
J0842$-$4851 & $-0.2$ & $0.7$ & $-$ & $-$ & $-$ & $-$ & J1602$-$5100 & $0.1$ & $0.7$ & $-$ & $-$ & $-$ & $-$ \\
J0904$-$7459 & $-5.0$ & $4.7$ & $-1.9$ & $0.9$ & $-3.8$ & $13.9$ & J1603$-$5657 & $1.7$ & $0.7$ & $0.8$ & $1.6$ & $2.1$ & $3.5$ \\
J0905$-$5127 & $0.1$ & $0.4$ & $-$ & $-$ & $-$ & $-$ & J1604$-$4909 & $0.6$ & $1.1$ & $1.7$ & $0.5$ & $0.0$ & $67.5$ \\
J0907$-$5157 & $-0.2$ & $0.2$ & $-0.2$ & $0.3$ & $-0.6$ & $0.3$ & J1605$-$5257 & $-2.0$ & $0.1$ & $0.0$ & $2.1$ & $-1.3$ & $0.6$ \\
J0908$-$1739 & $0.9$ & $0.5$ & $1.5$ & $2.4$ & $-0.2$ & $2.1$ & J1611$-$5209 & $-5.0$ & $5.3$ & $0.2$ & $2.6$ & $-3.5$ & $1.6$ \\
J0924$-$5814 & $-2.1$ & $1.0$ & $1.5$ & $0.5$ & $-0.7$ & $1.1$ & J1613$-$4714 & $-5.0$ & $7.8$ & $0.2$ & $0.7$ & $-0.7$ & $1.0$ \\
J0942$-$5552 & $0.8$ & $0.8$ & $0.0$ & $2.5$ & $-$ & $-$ & J1623$-$4256 & $0.2$ & $2.9$ & $0.1$ & $3.1$ & $-$ & $-$ \\
J0954$-$5430 & $-0.6$ & $0.3$ & $-1.0$ & $0.1$ & $-$ & $-$ & J1645$-$0317 & $5.0$ & $1.5$ & $1.7$ & $1.3$ & $0.6$ & $1.0$ \\
J1001$-$5507 & $1.0$ & $0.7$ & $0.2$ & $1.4$ & $-5.0$ & $12.3$ & J1646$-$6831 & $-1.0$ & $0.2$ & $-$ & $-$ & $-$ & $-$ \\
J1003$-$4747 & $0.4$ & $0.5$ & $1.2$ & $1.1$ & $0.6$ & $1.2$ & J1648$-$3256 & $-0.7$ & $1.1$ & $-4.2$ & $2.7$ & $3.9$ & $12.1$ \\
J1015$-$5719 & $-1.3$ & $0.1$ & $-0.1$ & $3.6$ & $-0.2$ & $1.8$ & J1651$-$4246 & $-5.0$ & $3.8$ & $0.1$ & $1.9$ & $-0.1$ & $2.2$ \\
J1017$-$5621 & $0.1$ & $8.9$ & $0.0$ & $2.5$ & $0.0$ & $22.9$ & J1651$-$5222 & $-0.1$ & $0.6$ & $-0.1$ & $1.4$ & $-1.4$ & $5.3$ \\
J1028$-$5819 & $0.5$ & $1.3$ & $-$ & $-$ & $-$ & $-$ & J1651$-$5255 & $1.5$ & $1.3$ & $-2.3$ & $2.0$ & $2.0$ & $0.9$ \\
J1038$-$5831 & $-0.2$ & $0.3$ & $-1.1$ & $0.4$ & $-0.2$ & $0.6$ & J1652$-$2404 & $-0.1$ & $1.7$ & $0.0$ & $6.9$ & $-0.1$ & $1.3$ \\
J1043$-$6116 & $-5.0$ & $7.9$ & $-2.4$ & $0.7$ & $-2.3$ & $0.8$ & J1703$-$3241 & $-1.1$ & $0.4$ & $-1.0$ & $0.4$ & $-0.1$ & $0.7$ \\
J1046$-$5813 & $0.1$ & $1.2$ & $0.1$ & $2.1$ & $0.0$ & $10.2$ & J1705$-$3423 & $-0.6$ & $0.2$ & $-1.8$ & $1.8$ & $-0.7$ & $0.4$ \\
J1048$-$5832 & $-0.6$ & $0.0$ & $-1.0$ & $0.6$ & $-0.3$ & $0.3$ & J1709$-$1640 & $-0.2$ & $0.1$ & $-1.6$ & $0.8$ & $-0.3$ & $0.2$ \\
J1049$-$5833 & $5.0$ & $3.4$ & $2.6$ & $2.1$ & $3.0$ & $1.9$ & J1709$-$4429 & $-1.2$ & $0.1$ & $-1.6$ & $0.2$ & $-2.4$ & $0.4$ \\
J1056$-$6258 & $2.1$ & $2.6$ & $-3.1$ & $1.1$ & $-2.9$ & $1.2$ & J1720$-$2933 & $-0.2$ & $0.7$ & $-0.2$ & $0.9$ & $-2.8$ & $1.3$ \\
J1110$-$5637 & $-2.0$ & $0.2$ & $-0.1$ & $0.8$ & $-4.2$ & $1.6$ & J1722$-$3207 & $-0.1$ & $0.9$ & $-0.3$ & $7.2$ & $-$ & $-$ \\
J1115$-$6052 & $-5.0$ & $9.6$ & $0.3$ & $5.1$ & $-5.0$ & $4.3$ & J1722$-$3632 & $-1.7$ & $0.3$ & $-0.1$ & $2.7$ & $-0.5$ & $3.3$ \\
\hline
\end{tabular}
\tablefoot{Columns are the best-fit power-law index and associated uncertainty from the three methods: total profile width at 20\% intensity ($\mu_{20}$), maximum Gaussian component separation ($\mu_{\rm sep}$), and individual component width ($\mu_{\rm width}$).}
\end{table*}

\begin{table*}
\centering
\small
\caption{Resulting frequency evolution parameters for profile widths (continued).}
\begin{tabular}{l|rr|rr|rr|l|rr|rr|rr}
\hline\hline
Pulsar Name & $\mu_{20}$ & $\sigma_{20}$ & $\mu_{\rm sep}$ & $\sigma_{\rm sep}$ & $\mu_{\rm width}$ & $\sigma_{\rm width}$ & 
Pulsar Name & $\mu_{20}$ & $\sigma_{20}$ & $\mu_{\rm sep}$ & $\sigma_{\rm sep}$ & $\mu_{\rm width}$ & $\sigma_{\rm width}$ \\
\hline
J1722$-$3712 & $5.0$ & $3.5$ & $-$ & $-$ & $-$ & $-$ & J1823$-$3106 & $-0.3$ & $0.1$ & $4.7$ & $5.0$ & $-0.2$ & $0.4$ \\
J1723$-$3659 & $-5.0$ & $2.9$ & $0.4$ & $8.0$ & $-1.6$ & $1.7$ & J1825$-$0935 & $1.8$ & $1.4$ & $0.0$ & $1.3$ & $-1.1$ & $0.4$ \\
J1727$-$2739 & $-1.8$ & $0.8$ & $-1.2$ & $0.2$ & $-0.1$ & $1.0$ & J1829$-$1751 & $-0.2$ & $0.2$ & $-4.2$ & $3.6$ & $-0.2$ & $1.3$ \\
J1731$-$4744 & $-0.8$ & $0.8$ & $0.0$ & $8.2$ & $-0.1$ & $3.0$ & J1830$-$1059 & $-0.5$ & $1.1$ & $-5.0$ & $2.2$ & $-0.5$ & $0.7$ \\
J1733$-$3716 & $-0.5$ & $0.2$ & $-0.1$ & $1.1$ & $-1.2$ & $0.9$ & J1832$-$0827 & $-2.4$ & $0.4$ & $4.2$ & $4.1$ & $-1.7$ & $0.7$ \\
J1735$-$0724 & $0.2$ & $1.9$ & $0.1$ & $4.1$ & $0.5$ & $3.4$ & J1834$-$0426 & $0.4$ & $0.4$ & $0.1$ & $1.9$ & $5.0$ & $7.0$ \\
J1738$-$3211 & $-5.0$ & $7.0$ & $-0.5$ & $0.9$ & $-$ & $-$ & J1835$-$1106 & $-2.4$ & $1.6$ & $-0.6$ & $1.5$ & $-3.3$ & $0.8$ \\
J1740$-$3015 & $-5.0$ & $3.0$ & $-5.0$ & $5.4$ & $-0.3$ & $0.9$ & J1842$-$0905 & $3.1$ & $2.8$ & $5.0$ & $16.3$ & $1.2$ & $7.4$ \\
J1741$-$3927 & $0.1$ & $2.1$ & $0.1$ & $0.8$ & $-0.1$ & $2.1$ & J1845$-$0743 & $3.9$ & $9.8$ & $-0.5$ & $0.3$ & $5.0$ & $6.9$ \\
J1745$-$3040 & $0.1$ & $0.6$ & $0.0$ & $1.2$ & $0.6$ & $0.1$ & J1847$-$0402 & $-0.4$ & $1.8$ & $-$ & $-$ & $-$ & $-$ \\
J1749$-$3002 & $0.0$ & $2.5$ & $0.0$ & $8.8$ & $-0.1$ & $3.4$ & J1848$-$0123 & $0.2$ & $1.0$ & $1.9$ & $8.4$ & $0.0$ & $7.9$ \\
J1750$-$3157 & $-4.7$ & $2.7$ & $-0.5$ & $0.4$ & $-$ & $-$ & J1852$-$0635 & $-0.9$ & $0.2$ & $-0.2$ & $0.2$ & $1.2$ & $10.1$ \\
J1751$-$3323 & $-4.4$ & $0.9$ & $0.4$ & $1.4$ & $-0.1$ & $2.0$ & J1900$-$2600 & $-0.1$ & $0.2$ & $-0.2$ & $1.3$ & $0.0$ & $10.4$ \\
J1751$-$4657 & $-0.1$ & $0.4$ & $-0.1$ & $0.6$ & $-0.1$ & $0.7$ & J1913$-$0440 & $0.2$ & $0.8$ & $0.3$ & $2.3$ & $-$ & $-$ \\
J1752$-$2806 & $-0.1$ & $2.5$ & $-0.5$ & $1.8$ & $-0.1$ & $2.8$ & J1932$+$2220 & $0.3$ & $0.8$ & $0.5$ & $2.2$ & $0.1$ & $2.9$ \\
J1803$-$2137 & $-0.1$ & $1.0$ & $-0.1$ & $7.0$ & $-0.2$ & $0.9$ & J1941$-$2602 & $2.9$ & $1.4$ & $0.0$ & $1.3$ & $0.2$ & $0.9$ \\
J1807$-$0847 & $0.0$ & $2.0$ & $-$ & $-$ & $-$ & $-$ & J2048$-$1616 & $-0.3$ & $0.2$ & $-$ & $-$ & $-$ & $-$ \\
J1817$-$3618 & $1.0$ & $2.9$ & $1.0$ & $0.9$ & $0.0$ & $5.9$ & J2330$-$2005 & $-0.1$ & $1.2$ & $0.0$ & $1.7$ & $0.0$ & $1.6$ \\
J1817$-$3837 & $0.4$ & $1.8$ & $4.4$ & $2.2$ & $-$ & $-$ & J2346$-$0609 & $-0.9$ & $0.3$ & $-0.3$ & $0.1$ & $-1.7$ & $2.4$ \\
J1820$-$0427 & $0.1$ & $1.4$ & $0.1$ & $1.3$ & $0.1$ & $3.2$ &  &  &  &  &  &  &  \\
J1822$-$2256 & $-5.0$ & $1.4$ & $-0.3$ & $1.0$ & $5.0$ & $5.1$ &  &  &  &  &  &  &  \\
\hline
\end{tabular}
\label{tab:pulsar_widths}
\end{table*}

\section{Results from Section 4.2.}

\begin{table*}
\centering
\caption{Resulting emission height behavior of individual pulsars. }
\begin{tabular}{lcc|lcc|lcc}
\hline
Pulsar & Group & Slope & Pulsar & Group & Slope & Pulsar & Group & Slope \\
\hline
J0034--0721 & 1 & $-$ & J1156--5707 & 1 & $-$ & J1707--4729 & $-$1 & $-$ \\
J0108--1431 & 1 & $+$ & J1157--6224 & 1 & $-$ & J1709--1640 & 0 & $+$ \\
J0151--0635 & 1 & $+$ & J1253--5820 & 1 & $+$ & J1709--4429 & 1 & $+$ \\
J0304+1932 & 0 & $-$ & J1302--6350 & $-$1 & $+$ & J1715--3903 & 1 & $+$ \\
J0401--7608 & $-$1 & $+$ & J1305--6203 & 1 & $-$ & J1715--4034 & $-$1 & $-$ \\
J0452--1759 & 0 & $-$ & J1306--6617 & $-$1 & $+$ & J1716--4005 & $-$1 & $-$ \\
J0536--7543 & $-$1 & $+$ & J1320--5359 & 1 & $-$ & J1717--3425 & $-$1 & $+$ \\
J0543+2329 & 1 & $-$ & J1326--5859 & $-$1 & $+$ & J1718--3825 & 1 & $+$ \\
J0601--0527 & $-$1 & $-$ & J1326--6700 & 0 & $+$ & J1720--2933 & 0 & $+$ \\
J0614+2229 & 1 & $-$ & J1327--6222 & $-$1 & $-$ & J1722--3632 & 1 & $-$ \\
J0624--0424 & $-$1 & $-$ & J1327--6301 & 1 & $-$ & J1722--3712 & 1 & $+$ \\
J0630--2834 & 1 & $+$ & J1328--4357 & 0 & $-$ & J1723--3659 & 1 & $-$ \\
J0631+1036 & 1 & $+$ & J1338--6204 & 1 & $+$ & J1727--2739 & $-$1 & $-$ \\
J0659+1414 & 1 & $-$ & J1352--6803 & 1 & $+$ & J1731--4744 & 0 & $+$ \\
J0729--1448 & 1 & $+$ & J1356--5521 & $-$1 & $+$ & J1733--3716 & 1 & $+$ \\
J0729--1836 & 1 & $+$ & J1357--62 & $-$1 & $-$ & J1735--0724 & $-$1 & $-$ \\
J0738--4042 & 1 & $+$ & J1357--6429 & 1 & $-$ & J1739--3023 & 1 & $-$ \\
J0742--2822 & 1 & $-$ & J1359--6038 & $-$1 & $+$ & J1741--2733 & $-$1 & $-$ \\
J0745--5353 & $-$1 & $+$ & J1401--6357 & 0 & $+$ & J1741--3016 & 1 & $-$ \\
J0758--1528 & 0 & $+$ & J1412--6145 & $-$1 & $+$ & J1741--3927 & $-$1 & $-$ \\
J0809--4753 & 1 & $-$ & J1420--6048 & 1 & $+$ & J1743--3150 & 1 & $+$ \\
J0820--1350 & 0 & $-$ & J1455--59 & 1 & $-$ & J1748--1300 & 1 & $+$ \\
J0820--4114 & 1 & $-$ & J1456--6843 & $-$1 & $+$ & J1749--3002 & 1 & $+$ \\
J0835--4510 & 1 & $-$ & J1513--5908 & $-$1 & $+$ & J1750--3157 & 1 & $+$ \\
J0842--4851 & $-$1 & $-$ & J1524--5625 & $-$1 & $-$ & J1751--3323 & $-$1 & $+$ \\
J0855--4644 & 1 & $+$ & J1524--5706 & $-$1 & $+$ & J1757--2421 & 0 & $+$ \\
J0857--4424 & 1 & $-$ & J1530--5327 & $-$1 & $+$ & J1801--2451 & $-$1 & $+$ \\
J0901--4624 & 1 & $-$ & J1531--5610 & $-$1 & $-$ & J1803--2137 & $-$1 & $+$ \\
J0904--7459 & 1 & $+$ & J1534--5405 & $-$1 & $-$ & J1807--0847 & 1 & $-$ \\
J0905--5127 & 1 & $-$ & J1535--4114 & 1 & $-$ & J1809--1917 & 0 & $+$ \\
J0907--5157 & $-$1 & $+$ & J1536--5433 & 0 & $+$ & J1816--2650 & 1 & $+$ \\
J0908--1739 & 1 & $-$ & J1539--5626 & $-$1 & $+$ & J1817--3618 & 1 & $-$ \\
J0924--5814 & 1 & $-$ & J1548--5607 & $-$1 & $+$ & J1822--2256 & 0 & $+$ \\
J0940--5428 & 0 & $-$ & J1559--4438 & $-$1 & $+$ & J1823--3106 & 1 & $-$ \\
J0942--5552 & 1 & $+$ & J1603--5657 & 0 & $+$ & J1826--1334 & $-$1 & $+$ \\
J0954--5430 & 0 & $-$ & J1605--5257 & 1 & $+$ & J1830--1059 & 1 & $-$ \\
J0959--4809 & 1 & $+$ & J1611--5209 & 0 & $-$ & J1833--0827 & $-$1 & $-$ \\
J1015--5719 & 1 & $+$ & J1637--4553 & 1 & $-$ & J1834--0426 & 1 & $+$ \\
J1016--5857 & $-$1 & $-$ & J1637--4642 & 1 & $-$ & J1835--1106 & 1 & $-$ \\
J1034--3224 & 1 & $+$ & J1643--4505 & $-$1 & $+$ & J1841--0345 & 1 & $-$ \\
J1038--5831 & $-$1 & $-$ & J1648--4611 & 1 & $-$ & J1842--0905 & 1 & $-$ \\
J1047--6709 & $-$1 & $+$ & J1649--4653 & $-$1 & $-$ & J1845--0434 & 1 & $+$ \\
J1048--5832 & 0 & $+$ & J1651--4246 & $-$1 & $-$ & J1845--0743 & 1 & $+$ \\
J1049--5833 & 1 & $+$ & J1651--5222 & 1 & $+$ & J1852--0635 & $-$1 & $-$ \\
J1056--6258 & 1 & $-$ & J1652--2404 & 1 & $-$ & J1853--0004 & 0 & $-$ \\
J1105--6107 & 1 & $-$ & J1653--3838 & $-$1 & $+$ & J1900--2600 & 1 & $+$ \\
J1110--5637 & $-$1 & $-$ & J1700--3312 & 1 & $+$ & J1910+0358 & $-$1 & $-$ \\
J1114--6100 & 1 & $+$ & J1701--3726 & 1 & $+$ & J1913--0440 & 1 & $-$ \\
J1115--6052 & 1 & $-$ & J1701--4533 & 1 & $+$ & J1932+2220 & 1 & $-$ \\
J1119--6127 & 1 & $-$ & J1702--4128 & $-$1 & $+$ & J1941--2602 & 1 & $+$ \\
J1123--6259 & 1 & $+$ & J1702--4310 & 1 & $-$ & J2048--1616 & $-$1 & $+$ \\
J1136--5525 & $-$1 & $+$ & J1703--3241 & 0 & $+$ & J2346--0609 & 0 & $-$ \\
J1146--6030 & 0 & $-$ & J1705--3950 & 1 & $+$ & & & \\
\hline
\end{tabular}
\label{tab:pulsar_classification}
\tablefoot{The table lists the classification based on the groups as classified by $\Delta \phi_0$ (column
2) and the correlation between emission height and observing frequency, with the slope indicating the trend: positive for increasing height with
frequency and negative for decreasing (column 3).}
\end{table*}

\end{appendix}

\end{document}